\documentclass[sigconf,authorversion]{acmart}

\usepackage[ruled]{algorithm2e} 
\usepackage{graphicx} 
\usepackage{subcaption} 
\usepackage{multirow} 

\definecolor{sfblue}{rgb}{0.1607843137254902,0.7098039215686275,0.9098039215686274}

\copyrightyear{2025}
\acmYear{2025}
\setcopyright{cc}
\setcctype{by}
\acmConference[ICPE '25]{Proceedings of the 16th ACM/SPEC International
Conference on Performance Engineering}{May 5--9, 2025}{Toronto, ON, Canada}
\acmBooktitle{Proceedings of the 16th ACM/SPEC International Conference on
Performance Engineering (ICPE '25), May 5--9, 2025, Toronto, ON, Canada}
\acmDOI{10.1145/3676151.3719353}
\acmISBN{979-8-4007-1073-5/25/05}

\makeatletter
\def\@affiliationfont{\normalsize\normalfont}
\makeatother

\begin{document}


\title{Shaved Ice: Optimal Compute Resource Commitments for Dynamic Multi-Cloud Workloads}
\settopmatter{authorsperrow=4} 

\author{Murray Stokely}
\orcid{0009-0008-3390-1338}
\affiliation{%
\institution{Snowflake, Inc.}
\city{San Mateo}
\state{CA}
\country{USA}
}
\email{murray.stokely@snowflake.com}

\author{Neel Nadgir}
\orcid{0009-0009-4283-0137}
\affiliation{%
\institution{Snowflake, Inc.}
\city{San Mateo}
\state{CA}
\country{USA}
}
\email{neel.nadgir@snowflake.com}

\author{Jack Peele}
\orcid{0009-0002-0487-3709}
\affiliation{
 \institution{Snowflake, Inc.}
  \city{San Mateo}
  \state{CA}
  \country{USA}
}
\email{jack.peele@snowflake.com}
\author{Orestis Kostakis}
\orcid{0009-0006-3476-0886}
\affiliation{%
\institution{Snowflake, Inc.}
\city{Bellevue}
\state{WA}
\country{USA}
}
\email{orestis.kostakis@snowflake.com}

\begin{abstract}


Cloud providers have introduced pricing models to incentivize long-term commitments of compute capacity. 
These long-term commitments allow the cloud providers to get guaranteed revenue for their investments in data centers and computing infrastructure.
However, these commitments expose cloud customers to \emph{demand risk} if expected future demand does not materialize.
While there are existing studies of theoretical techniques for optimizing performance, latency, and cost, relatively little has been reported so far on the trade-offs between cost savings and demand risk for compute commitments for large-scale cloud services.

We characterize cloud compute demand based on an extensive three year study of the Snowflake Data Cloud,
which includes data warehousing, data lakes, data science, data engineering, and other workloads across multiple clouds.
We quantify capacity demand drivers from user workloads, hardware generational improvements, and software performance improvements.
Using this data, we formulate a series of
practical optimizations that maximize capacity availability and minimize costs for the cloud customer.
\end{abstract}


\maketitle

%



\section{Introduction}
\label{sec:intro}
An increasing percentage of the world's compute workloads are moving to cloud infrastructure built on top of compute, storage, and networking resources provided by Cloud Service Providers (CSPs), such as Amazon AWS \cite{aws}, Microsoft Azure \cite{azure}, and Google GCP \cite{gcp}.  These CSPs invest billions of dollars building out physical data center infrastructure, and then rent these resources at different price and performance tiers to allow platform and application owners to balance cost against specific thresholds of availability, latency, and performance.  The CSPs also offer different prices based on the level of commitment and specificity of the demand.  For example, there may be one price to use a Virtual Machine (VM) with a fixed set of resources for an hour, but a much lower price per hour to commit to using that same VM continuously for a set longer period of time.

These considerations provide economic incentives for consumers of cloud resources to lock-in capacity commitments from CSPs that provide greater discounts to meet the required demand.  However, accurately forecasting future demand
 is
 challenging due to growth and changes in user demand, new product features with different resource requirements, and availability of new generations of
hardware with different capabilities from the CSPs.  This fundamental mismatch between the drivers of supply and demand between CSPs and consumers of cloud resources motivates
several
opportunities for optimization.  We are not aware of existing work that has considered real-world data of the demand changes from large-scale deployments nor the performance changes from the CSPs.

Our work is motivated by the following research question: \emph{Can we quantify the tradeoffs between the economic incentives of making longer capacity commitments against the risk of demand changes leading to surplus unused capacity in a large Multi-Cloud service?}  We make seven contributions to help answer this question:

\begin{enumerate}
\item We collect and process over three years of VM demand from one of the world's largest Multi-Cloud data platforms, and release an anonymized subset of this dataset. (\S~\ref{sec:dataset}) 
\item We analyze the patterns of user demand at different time granularities, showing daily, weekly, and annual cycles. (\S~\ref{sec:userdemand})
\item We analyze the performance of multiple generations of compute VMs available for user workloads over three years from three different CSPs. (\S~\ref{sec:hardwaredemand})
\item We analyze the improvements of software performance on stable workloads over that time period. (\S~\ref{sec:softwaredemand})
\item Using this analysis from (2), (3), and (4), we formulate an optimization problem for long-term capacity commitments to minimize cost and quantify risk. (\S~\ref{sec:commitment})
\item We analyze Snowflake workloads to identify \emph{time shifting} candidates to move from peak to off-peak times to further take advantage of the optimal commitment level. (\S~\ref{sec:timeshifting})
\item We characterize performance of three CSPs at providing VMs at scale and formulate a second optimization problem to minimize costs for maintaining a minimal pool of pre-provisioned VMs to improve performance. (\S~\ref{sec:freepool})
\end{enumerate}


\begin{figure*}[t!]
  \includegraphics[width=\textwidth]{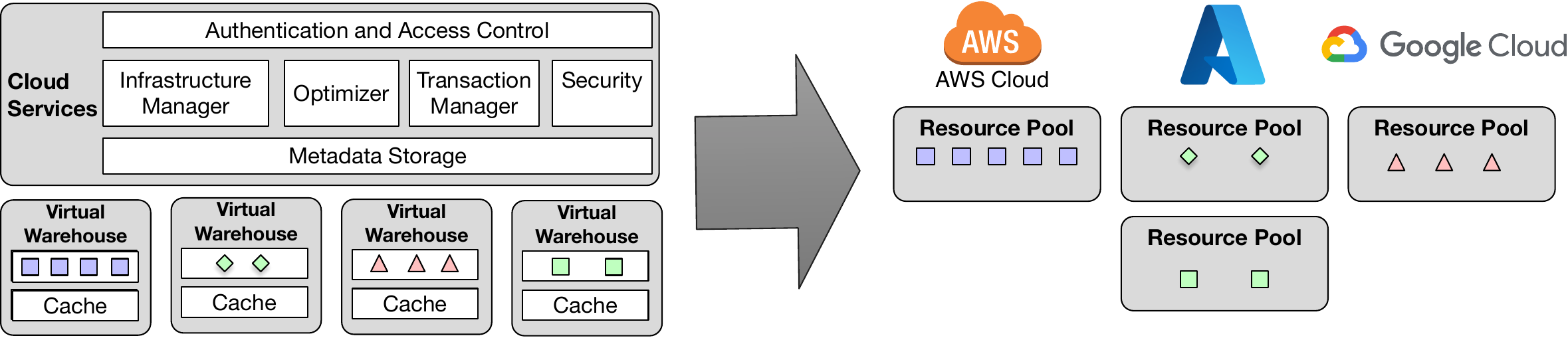}
  \caption{Multi-Cluster, Shared Data Architecture of Snowflake (left) Mapped to Individual Resource Pools of Different Shaped Compute VMs on Different Infrastructure-as-a-Service (IaaS) Cloud Providers (right).}
  \Description[Architecture diagram of Snowflake system components map to resource pools on different cloud providers]{The major components of the Snowflake architecture are Cloud Services layer and Virtual Warehouses.  These software components fundamentally map to different resource pools in the Cloud Service Providers, for example, a specific VM type.}
  \label{fig:arch}
\end{figure*}





\section{Characterizing Cloud Compute Demand Drivers}
\label{sec:workloads}

Demand for underlying CSP resources from an application is a function of several different demand drivers, including:
\begin{itemize}
\item The user demand for the application, often expressed as number of requests or queries.
\item The software efficiency of that application, measured as the resources consumed per request or query.
\item The hardware capabilities of each capacity unit provided by the CSP.
\item The utilization of the underlying CSP resources by the application.
\end{itemize}

In this section we characterize these demand drivers before we optimize capacity commitments for this demand in \S~\ref{sec:commitment}.

\subsection{The Snowflake Workload}

Snowflake \cite{snowflake} is a multi-tenant cloud data warehouse offering a
highly available, scalable and elastic database as a service. Snowflake
has a multi-cluster, shared-data architecture and adopts a service-oriented design where both compute and storage layers are independently scalable.  It is used by customers around the world to answer more than 6 billion queries per day over exabytes of data.

The two major layers of the compute architecture of Snowflake are:

\begin{itemize}
\item \textbf{Cloud Services} \cite{snowflakecp} The collection of services that comprise the control plane. They manage virtual warehouses to respond to user queries, transactions, and all of the associated metadata.
\item \textbf{Virtual Warehouses} A scalable cluster of worker nodes for query execution running on virtual machines.
\end{itemize}

Figure~\ref{fig:arch} illustrates these two compute layers and how they are mapped into resource pools of different VM sizes on the different cloud providers.  Each resource pool represents a uniform compute SKU with a set hardware generation and ratio of compute, memory, storage, and networking resources, and the system autoscales \cite{autoscaling} up and down the appropriate resource pool depending on the workload.



\subsection{User Demand Patterns}
\label{sec:userdemand}

Elastically scaling up virtual warehouses allows users to match resource consumption to their immediate needs, but how do those needs typically vary over longer periods of time?  Workloads from Spark jobs \cite{databricksworkload} and enterprise data warehouses \cite{facebookdw} have previously been shown to include significant periodicity, and Snowflake is no different.

Vuppalapati, et al. \cite{snowflakeelasticquery} released a data set with execution details of 70 million Snowflake queries. 
The read-only ad-hoc and interactive OLAP queries were characterized as having
significant periodicity corresponding to daytime hours on weekdays.
Looking at the total query volume across all query types between February 22 and March 7, 2018 shows that on an average day, 35\% more queries were executed at the peak hour compared to the minimum hour.  A similar pattern is still visible 6 years later in the number of VM instance-hours used by Snowflake over time.

\begin{figure}[t!]
  \centering
  \includegraphics[width=\linewidth]{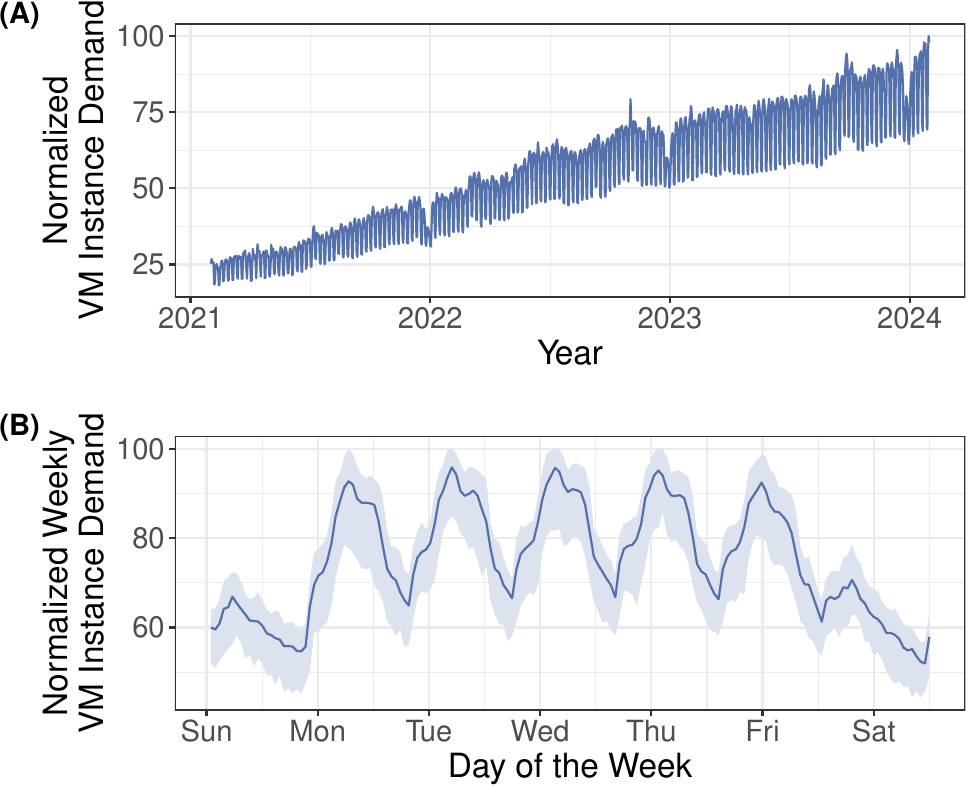}
  \caption{(A) Daily demand of VM instances in a Snowflake compute resource pool over 3 years showing significant seasonality with an end of year drop in demand, and (B) Median hourly VM instance demand over a week showing significant daily and weekly periodicity.  The confidence interval describes the 95th percentile for all weeks in the full 3-year history.}
  \label{fig:dailypattern}
  \Description{The top plot, labeled A, shows VM instance demand (y-axis) over time (x-axis) from early 2021 through early 2024 with a steady upward trend, weekly cycle, and three small reductions in demand around the end of each calendar year.  The bottom plot, labeled B, shows hourly VM instance demand (y-axis) over each of the 7 days of a week, showing a clear daily pattern of higher demand during business hours and a reduction in demand on nights and weekends.  This includes a confidence interval highlighting the range of all weeks in the 3 year dat set.}
\end{figure}

Figure~\ref{fig:dailypattern} shows the number of VM instance-hours from an example resource pool used to serve user queries over a three-year period.
Two plots are shown to highlight the periodicity at two different time scales.
The top plot shows a significant seasonal drop in demand at the end of each calendar year, and the bottom plot shows the weekly pattern with higher usage during business hours and reduced usage over the weekends.

For the full 3-year time series at daily granularity, the lag-7 autocorrelation is 0.975, and the average weekly maximum is 35\% higher than the average weekly minimum, a weekly effect.  At hourly granularity, the average daily maximum is 34\% higher than the average daily minimum, a diurnal effect.  This daily effect is higher than the daily variation in CPU utilization reported in Meta (20\%) and Google (15\%) data centers found in \cite{brookstimeshifting}.\footnote{The Meta and Google data center numbers represent utilization of a fixed number of servers throughout the day, while the Snowflake number represents the variation in the number of VM instances that are autoscaled throughout the day with user demand.  Although not exactly measuring the same thing, we believe this is a fair way to report the impact of diurnal traffic patterns on fixed data center infrastructure vs tenants on Cloud VMs that can simply give back the VMs off-peak.}


\subsection{Hardware Evolution Demand Impact}
\label{sec:hardwaredemand}

In addition to changes in demand from user workloads, we must also consider the impact of new hardware generations offered by the CSPs.  Barroso \cite{priceperformance} showed significant price-performance improvement over three generations of Google servers, and Armbrust, et al. \cite{abovetheclouds} showed a 16x improvement in price-performance from 2003 to 2008.  If price-performance of compute offerings provided by the CSPs is growing faster than user demand, this may impact the types of long-term capacity commitments that can be profitably utilized.

Each of the three major CSPs considered here have introduced a number of platform changes to their VM offerings in recent years:
\begin{enumerate}
\item Introduction of Arm-based processors
\item Transition to DDR5 memory
\item Improvements in the performance of local SSD
\item Increased networking bandwidth
\end{enumerate}

The combination of Arm and DDR5 memory has provided up to a 50\% improvement in memory bandwidth.
Memory
bandwidth is critical for the Hash Table and Bloom Filter implementations of our workload, and
this increase in bandwidth directly translates to lower query latency.
These VM
instances also have increased network bandwidth which helps in data
transfer between nodes (e.g.\ during distributed hash joins).
Improvements in
the performance of the local SSD translates to faster access time to ephemeral
storage.

Using a combination of industry standard benchmarks like
TPC-DS \cite{tpcds} and replaying actual customer queries using
Snowtrail \cite{snowtrail}, we have measured improvements in
query latency across major hardware transitions over the past three years.
Graviton 3 instances provide Snowflake customers a 25\% median reduction in query
latency compared to the previous generation.
This matches query latency reductions measured by AWS with popular Open
Source databases (19\% with PostgreSQL, 29\% with MySQL, 34\% with
MariaDB) \cite{awsg3perf1}.
Table~\ref{tab:hwperfimprove} shows this and three other examples of hardware generation
transitions that created step-functions in performance that were material enough
to impact long-term capacity commitments \cite{awsg3perf,awsg4perf,gcpperf}.

Individual queries in this workload have different capacity bottlenecks on CPU, I/O, memory, and other resources.
These bottlenecks evolve over time, and different hardware transitions may disproportionately benefit certain parts of the workload.
Previous analysis \cite{snowflakeelasticquery} found, for example, that the amount of data read can vary over nine orders of magnitude for queries.
This illustrates how a network-bound query may not see gains with a faster CPU.
While a wider range of SKUs offers flexibility for the diverse needs of different query types, it also increases testing
complexity and reduces economies of scale, making it desirable to find a smaller
subset that meets most customer needs \cite{softsku}.

Having a common pool of instance types
also enables workload decorrelation, which reduces the spikes in resource demand.
Since virtual warehouses map to a small number of resource pools of different VM instance types provided by the CSPs,
the aggregate impact of different types of queries can be measured and used as an input to the capacity planning process.
The large aggregate step-function increases in performance are critical to incorporate into the optimization process for deciding long-term commitment levels described in \S~\ref{sec:commitment}.

\subsection{Software Performance Improvements}
\label{sec:softwaredemand}

\begin{table}
  \caption{Reduction in Median Latency on SnowTrail Queries Provided by VM Hardware Platform Changes.}
  \label{tab:hwperfimprove}
  \begin{tabular}{ccccl}
    \toprule
    Date & CSP    & Old Platform  & New Platform  & \% Reduction \\
    \midrule
    05/2022 & AWS    & Graviton2     & Graviton3     & 25\%\\
    08/2024 & AWS    & Graviton3     & Graviton4     & 30\%\\
    09/2022 & Azure  & DPv5          & DPv6          & 20\%\\
    04/2024 & GCP    & X86           & Axion         & 50\%\\
    \bottomrule
  \end{tabular}
\end{table}

The computing industry has relied on automatic performance increases of Moore's Law for decades, but software code efficiency improvements through algorithmic changes and low-level performance optimizations to utilize hardware more effectively are becoming increasingly important for reducing cloud capacity costs \cite{codeefficiency}.  For the Snowflake workloads introduced in this section, we have seen performance improve by 12\% in the last year \cite{snowspi}.
These improvements are a combination of systemic and
algorithmic improvements. We introduced a highly parallelizable encryption
mechanism that effectively doubles the network bandwidth and resulted in
bandwidth-heavy queries executing up to 40\% faster \cite{snowadaptnet}. Similarly, we
optimized Top-K pruning, which benefits queries that use \texttt{ORDER} and
\texttt{LIMIT} clauses by an average of 12.5\% \cite{spi27blog}.
Our
optimizations in memory management for holistic and adaptive broadcast join
decisions \cite{spi27blog} improve queries with joins.

We have also rolled out and tested more general performance improvements such as compiler upgrades and profile guided optimizations.
As with the hardware improvements that benefit different parts of the workload differently, the fact that virtual warehouses map to a small number of resource pools of different VM instance types provided by the CSPs allows us to measure the aggregate impact of these performance changes as an input to the capacity optimization formulated in the next section.

\section{Commitment Levels}
\label{sec:commitment}

\subsection{Definitions}

The three most prominent pricing options that are widely supported \cite{reducecostsbrazil} by leading IaaS clouds for Virtual Machine (VM) capacity are:

\begin{enumerate}
\item \emph{On-Demand} instances where users pay only for the incurred instance-hours, without making any long-term commitment.
\item \emph{Reserved Instances or Savings Plans} where users reserve capacity for months or years in exchange for a significant discount compared to the hourly on-demand rate.
\item \emph{Spot} instances that can be interrupted by the IaaS provider and rented via an auction system at lower prices.
\end{enumerate}

We focus here on the first two categories, because availability of spot instances may be limited at large scale, and the interruptibility is not suitable for all workloads.
Table~\ref{tab:savingsplan} shows the discount for 3-year capacity commitments relative to the base price for On-Demand capacity across 8 representative VM instance families from 3 different public clouds.
The pricing data \cite{awsprice,azureprice,gcpprice} shows that On Demand capacity is on average 2.1x more expensive than 3-year savings plan commitments.

\begin{table}
  \caption{Savings Plan Discounts from Cloud Providers.}
  \label{tab:savingsplan}
  \begin{tabular}{cccc}
    \toprule
\multicolumn{1}{l}{Cloud} & \multicolumn{1}{l}{Instance Family} & \multicolumn{2}{c}{Savings Plan Discount} \\
                          &                                     & 1 year & 3 year             \\
    \midrule
AWS                       & C6i                                 & 28\%                & 52\%               \\
AWS                       & C7i                                 & 28\%                & 52\%               \\
AWS                       & C7GD                                & 28\%                & 52\%               \\
AWS                        & M7GD                               & 27\%                & 50\%                \\
Azure                     & Std\_Dd\_v4                         & 31\%                & 54\%               \\
Azure                     & Std\_Dpd\_v5                        & 31\%                & 54\%               \\
GCP                       & N2-Standard                         & 37\%                & 55\%               \\
GCP                       & N4-Standard                         & 37\%          & 55\% \\
    \bottomrule
\end{tabular}
\end{table}

\begin{figure}
 \includegraphics[width=\linewidth]{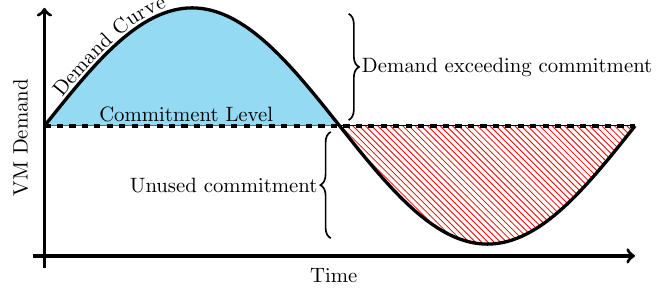}

    \caption{Example demand curve with fixed capacity commitment showing area of demand exceeding commitment level (blue) and area of unused commitment (red hatches).}
    \Description[Example demand curve]{Example demand curve with time as the x-axis and VM demand as the y-axis showing a fixed horizontal line of the commitment level.  The example demand curve is represented as a sine wave with the area above the commitment level and below the demand curve shaded in blue and labeled as demand exceeding commitment, and the area below the commitment level and above the demand curve at different time intervals shaded with red hatches and labeled as unused commitment.}
    \label{fig:exdemand}
\end{figure}

The periodic demand we characterized in the previous section provides a challenge for fixed capacity commitment levels.  Figure~\ref{fig:exdemand} shows an example periodic demand function, $f(x) = sin(x)$, with a fixed capacity commitment level highlighting the different costs for demand exceeding the commitment level and the unused commitment.
The area of demand exceeding the commitment level incurs an on-demand price premium, and the area below the commitment level represents unused or forfeited commitment.

\subsection{Setting Optimal Commitment Level for Periodic Demand}
\label{sec:optimallevel}

Techniques such as \cite{prosperopscyclical} have been proposed to maximize savings with cyclical demand, but these approaches have significant open questions:  What is the best time window to consider for computing the optimal savings plan level?  How can forecasts of future organic growth be incorporated?  How do known feature launches and hardware migrations affect this?  What is the impact of holidays and how should purchase strategies shift?

For a given demand curve $f(x)$ of the number or cost of VM instances needed over time (such as Figure~\ref{fig:dailypattern}) we seek to identify the horizontal line $y=c$, where $c \in (\mathrm{min}(f(x)), \mathrm{max}(f(x)))$, such that the total weighted cost of the area between the horizontal line and the demand curve above and below that line is minimized.

More formally, we seek to minimize the following cost function $C(c)$:

\begin{equation} \label{eq_cost}
\begin{split}
C(c) = & \int_0^T A \cdot (f(x) - c) dx \mathrm{~for~} f(x) > c\\
   & +  \int_0^T B \cdot (c-f(x))dx \mathrm{~for~} f(x) < c
\end{split}
\end{equation}

where:
\begin{enumerate}
\item $A$ is the cost factor for the area above the line (e.g. On-Demand Capacity)
\item $B$ is the cost factor for the area below the line (e.g. Unused Savings Plans)
\end{enumerate}

In general $f(x)$ is an empirical curve of real-time demand for cloud services and so our cost function $C(c)$ is not necessarily an integrable function where the minimum can be found analytically.  However, because $f(x)$ will in practice be measured at discrete time units we can use numeric techniques such as 
Brent's Method \cite{brent1973algorithms}.

\begin{figure}
  \centering
  \includegraphics[width=\linewidth]{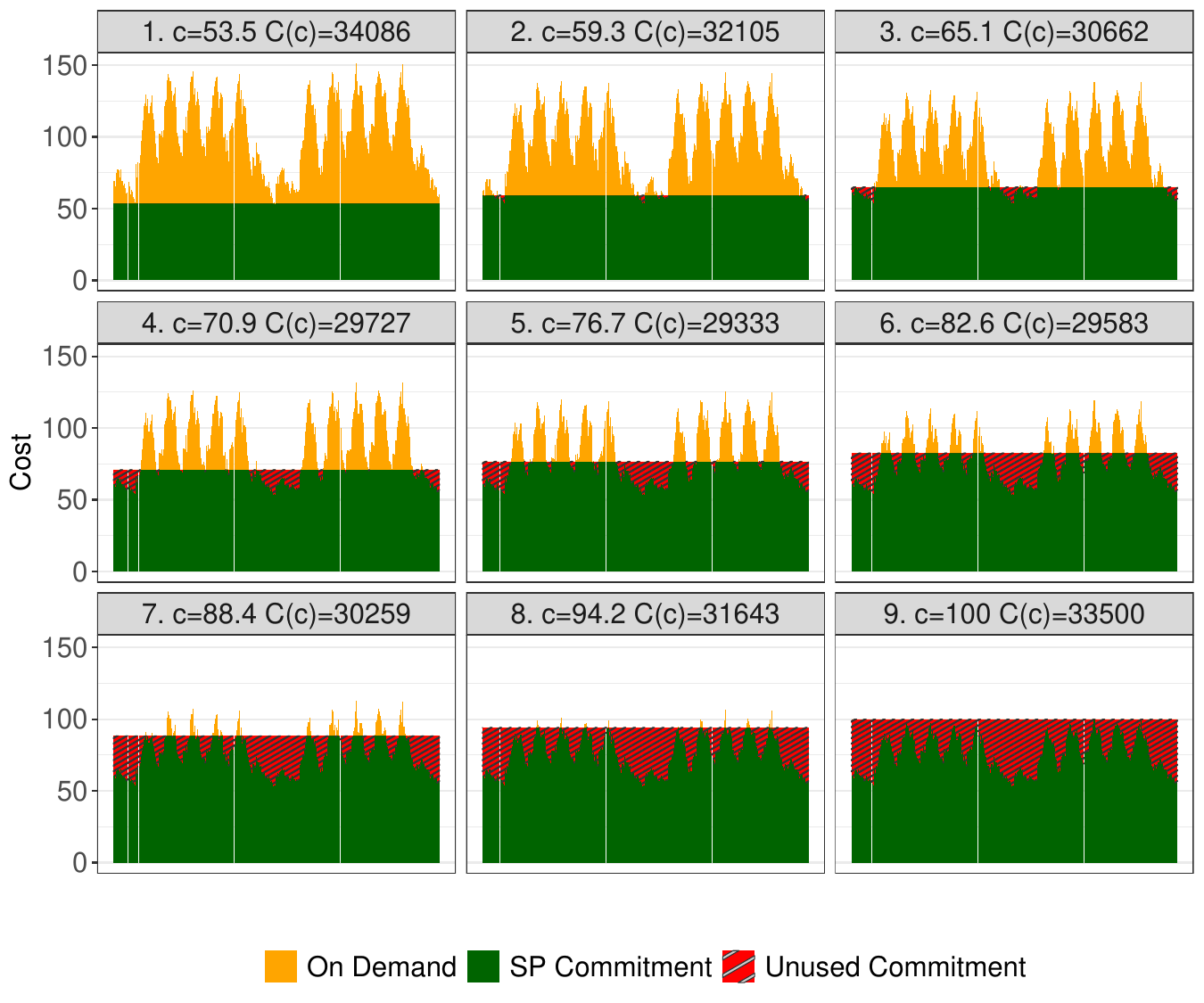}
  \caption{Visualization of the numeric approximation method for computing the commitment-level $c$ that has minimal cost, $C(c)$, for an empirical demand curve with daily and weekly periodicity and weights $A = 2.1, B=1.0$.}
  \label{fig:threeby}
  \Description{A matrix of 9 plots arranged in a 3x3 grid where each plot shows the same two-week period of VM instance demand under different savings commitment level, c.  The x-axis of each plot is time and the Y-axis of each plot is total VM demand cost.  The upper left plot starts with the commitment level c set to the minimum demand throughout the two-week period, and the bottom right plot ends with the commitment level c set to the maximum demand throughout the two-week period.  Each of the intermediate plots represents a step between the two extremes.  The graph is colored to indicate the area of demand covered by the capacity commitment (in green), the demand that must be covered at on-demand rates (in yellow), and the unused capacity commitment (in red with hatches).  Each plot is labeled with the value of c and the cost function for that value C(c) and it is visually shown that the minimal cost strategy is in figure 5, between the two extremes.  Scenario 1 shows c=53.5 and C(c)=34086, scenario 2 shows c=59.3 and C(c)=32105, scenario 3 shows c=65.1 and C(c)=30662, scenario 4 shows c=70.9 and C(c)=29727, scenario 5 shows c=76.7 and C(c)=29333, scenario 6 shows c=82.6 and C(c)=29583, scenario 7 shows c=88.4 and C(c)=30259, scenario 8 shows c=94.2 and C(c)=31643, and scenario 9 shows c=100 and C(c)=33500.  The minimum cost option from these 9 scenarios is scenario 5 when c=76.7 and C(c)=29333 and the scenarios with lower and higher commitment levels to this both have higher cost.}
\end{figure}

Figure~\ref{fig:threeby} shows an illustration of the total cost of cloud compute resources under 9 different scenarios of commitment levels set given the periodic workload observed in Snowflake over a two-week period of time in January 2024.  In this example, we assume our capacity commitment has a unit cost whether it is used or not.
We take the mean Savings Plan discounts reported in Table~\ref{tab:savingsplan} and use $2.1$ as our cost factor for on-demand capacity above our commitment level.

Each of the 9 different scenarios shows the commitment level as a flat line, with demand exceeding the commitment level shown in yellow, fulfilled demand in green, and unused capacity when demand is below the commitment level shown in red with diagonal hatches.  If we set the commitment level to the minimum amount of demand over our time period, there is no unused capacity commitment.  If we set the commitment level to the maximum amount of demand over our time period, there is no expensive on-demand cost.  By computing the cost, $C(c)$ in each scenario we see that scenario 5 provides the lowest cost option for these specific parameters.

\subsection{Minimizing Risk of Longer-Term Commitments}

The last section described how to compute the optimal commitment level over an empirical data set, but these commitments must be made in advance when there is imperfect knowledge about future demand.  To illustrate the challenges of this method, we analyzed three years of history to measure trend, seasonal holiday effects, and inorganic changes in demand.

\subsubsection{Trend}

\begin{figure}
  \centering
  \includegraphics[width=\linewidth]{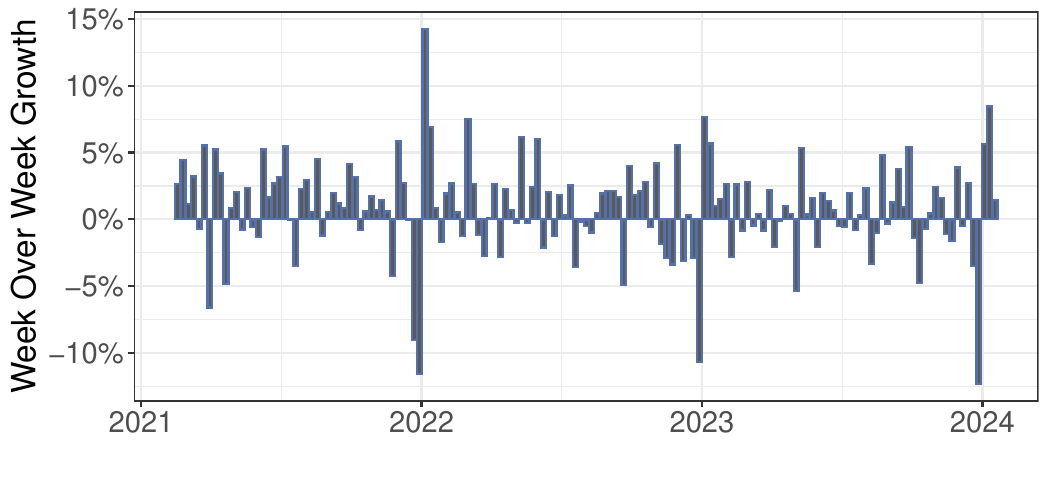}
  \caption{The week over week growth rate of aggregate VM demand in the dataset shows a significant number of negative weeks, including a clear seasonal pattern, despite a high annual growth rate.}
  \label{fig:wowtrend}
  \Description{A graph showing the week over week growth rate (y-axis) over time for each week from early 2021 to early 2024.  The y-axis is labeled from -10\% to +15\% and shows weekly growth rates in the range of -7\% to +8\% for each week with five exceptions.  The last week of each of the three years shown corresponds to a large negative week over week decline exceeding -10\%.  The second to last year of 2021 also has a larger than usual -8\% decline in demand, and the first week of 2023 has a higher than normal 14\% increase in demand.}
\end{figure}

We start by considering trend. 
 If trend can be estimated then the future forward-looking capacity commitments can be made more accurately to reduce cost.  In Figure~\ref{fig:dailypattern} we see that the three-year timeseries of VM demand for the resource pool shown is 3.9 times higher at the end of the period than the beginning, corresponding to a 58\% annual growth rate.  A positive trend like this is important for entering into long-term capacity commitments since these commitments can not normally be reduced in size.  But what about on smaller timescales?  Are there more localized reductions in trend that would motivate us to pause the purchase of additional savings plan commitments to minimize cost?  Figure~\ref{fig:wowtrend} shows the week over week growth rates in mean VM demand.  Despite the significant overall growth rate in VM demand, the average week over week trend is negative 37\% of the time.  If we can forecast in advance periods of time where there will be a reduction in VM demand we can adjust long-term commitment purchases to avoid over-commitment before a drop in demand.

\begin{figure}
  \centering
  \includegraphics[width=\linewidth]{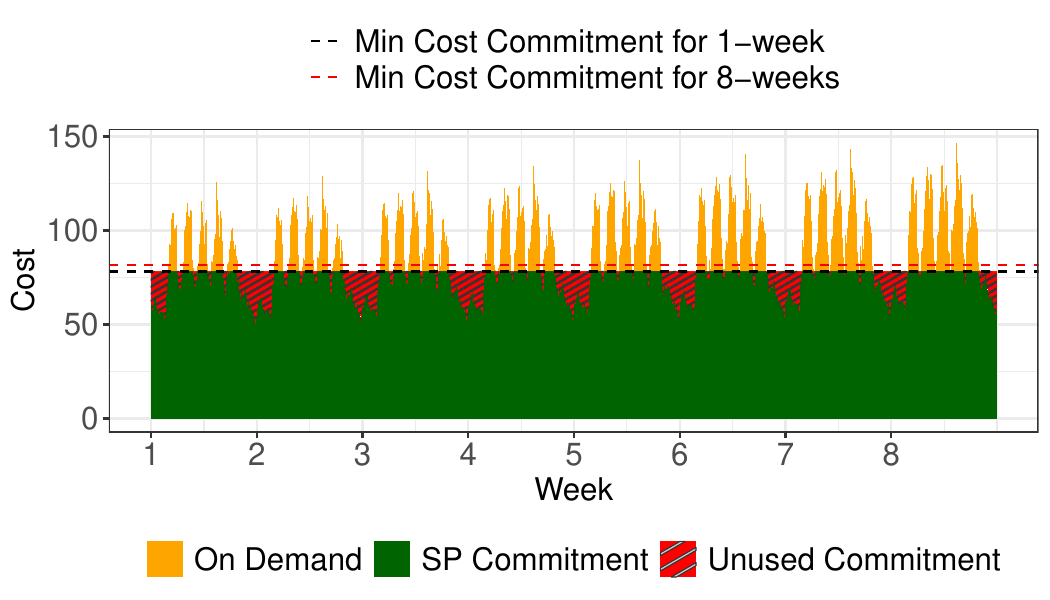}
  \caption{One week of Snowflake workload extrapolated out eight weeks with a simulated annual trend of 100\%.  The dashed black line shows the optimum commitment level when considering only the first week.  The dashed red line shows the optimum commitment level when considering the full eight-week period.  When commitments can be increased but not decreased it is important to consider upcoming changes in trend.}
  \label{fig:trendforecast}
  \Description{A graph showing total VM demand cost (y-axis) over eight weeks of total VM demand cost.  The clear weekly pattern is continued either weeks into the future by adding a simulated 100\% annual trend so that each week has slightly higher peaks than the week prior.}
\end{figure}

%
%
Figure~\ref{fig:trendforecast} shows the last full week of our example data set extended out eight weeks with a simulated annual trend of 100\%.
The two dashed lines highlight two different commitment levels that are set by using different forward-looking forecast horizons for the optimization.
More generally, Table~\ref{tab:trendsensitivity} compares the cost differences between using actual demand and forecasts with varying horizons and annual trend rates.
This table shows that when new capacity commitments are made each week or more frequently, the trend has limited impact for additional savings.
When capacity commitments are made less frequently, however, it is important to use the trend in order to unlock additional savings.


\begin{table}
\caption{Sensitivity analysis of $C_1(c) - C_2(c)$ cost delta per million dollars when commitment level computed from actuals, $C_1$, vs computed from forecast, $C_2$, with the specified annual trend and forecast horizon.  With a higher annual trend and longer forecast horizon we see significantly larger cost increases.}

\label{tab:trendsensitivity}
\centering
\begin{tabular}{cl|r|r|r|r|r}
& & \multicolumn{5}{c}{Annual Trend}\\
& & 10\% & 25\% & 50\% & 75\% & 100\%\\
\toprule
\parbox[t]{2mm}{\multirow{3}{*}{\rotatebox[origin=c]{90}{Update Freq}}} & 1 week & \$7.05 & \$16.50 & \$29.04 & \$35.84 & \$41.71\\
& 2 week & \$3.51 & \$8.21 & \$67.21 & \$137.12 & \$207.48\\
& 4 week & \$1.75 & \$60.32 & \$237.99 & \$512.68 & \$866.13\\
& 8 week & \$32.62 & \$277.17  & \$1174.65 & \$2321.38 & \$3612.77\\
\bottomrule
\end{tabular}
\end{table}

\subsubsection{Seasonality}

So far we have considered daily and weekly periodicity as well as long-term trend in our demand functions, but what about seasonal variations?  If we ignore major seasonal effects we may significantly over or under-estimate the optimal level of capacity commitments to purchase.

\begin{figure}
  \centering
  \includegraphics[width=\linewidth]{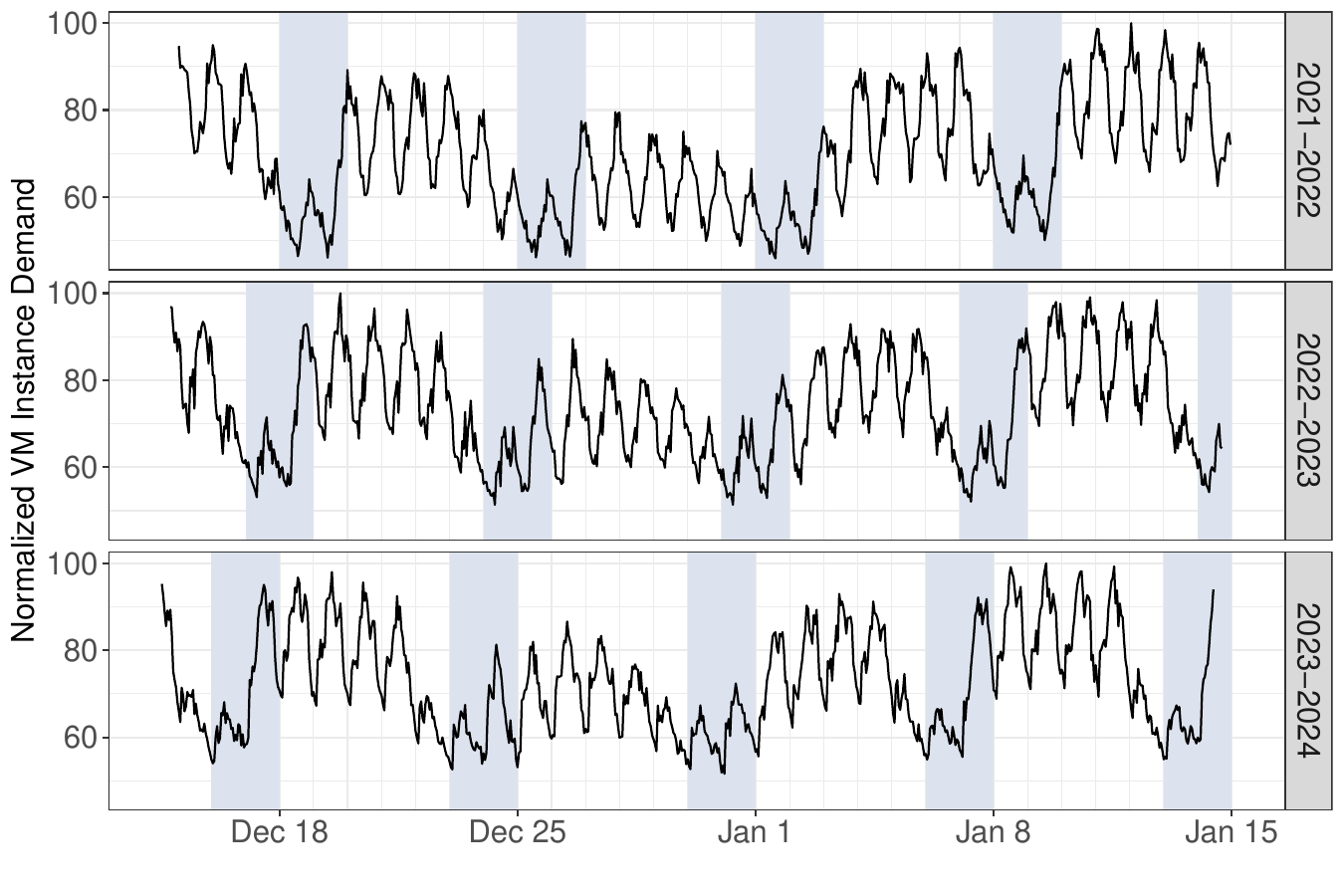}
  \caption{Hourly VM instance demand between December 15 and January 15 over three successive years.  Each time series panel is normalized by the maximum hourly demand in that panel.   Weekends are highlighted and there is a noticeable seasonal demand drop between Christmas and New Years.}
  \label{fig:christmas}
  \Description{Three graphs of hourly VM instance demand cost between December 15 and January 15 over three successive years: 2021-2022, 2022-2023, and 2023-2024.  The key weekend and holiday characteristics are explained in the caption.}
\end{figure}

Figure~\ref{fig:christmas} shows a subset of Snowflake cloud compute usage over the period from December 15 until January 15 in each of three consecutive years.
The two-week period around Christmas and New Years holidays had on average 8\% less demand than the preceding two-week period of time.  The forward-looking window of time under which the optimization from \S~\ref{sec:optimallevel} is run must be large enough to set the commitment level taking into account this large drop.

\subsubsection{Forecasts}

Long-term commitments can be increased by purchasing additional incremental commitments, but cannot normally be reduced if the level is set too high.  For this reason, forecasting future demand taking into account all of the demand drivers of the previous sections and then purchasing optimal capacity commitments for that demand is important to minimize overall cost.  Algorithm~\ref{alg:forecast} illustrates how we incorporate forecasts of future demand into our optimization.


We use a multi-year training dataset to generate forecasts of future demand up to 1 year in the future using the Prophet \cite{prophet} forecast model tuned with a weighted error metric\footnote{Our error costs are asymmetric since the cost of under-forecasting and paying On-Demand rates is $2.1\times$ more expensive compared to the cost of over-forecasting} (\emph{Step 1}).  We then run the optimization formulation from the last section over different forecast horizons (\emph{Steps 2 and 3}), and then take the minimum of these optimization outputs as our recommended compute capacity commitment (\emph{Step 4}).

\begin{algorithm}
\KwIn{An hourly timeseries training dataset $X_t$ for $t = 1, 2, \ldots, T$ where $T$ is the length of the historic training dataset.}
\KwOut{The minimum cost commitment level, $c^*$, to purchase now, assuming we can purchase additional incremental commitments in the future but can not reduce the commitment.}
\textbf{Step 1:} Generate an hourly forecast with a one year forecast horizon, $\hat{X}_{T+52 \times 24 \times 7}$.

\textbf{Step 2:} For each weekly forecast horizon $w = 1, 2, \ldots, 52,$ extract the subset of forecasted hourly data up to $w$ weeks in the future.

\begin{displaymath}
\hat{X}_{w_n} = { \hat{X}_{T+1}, \hat{X}_{T+2}, \ldots, \hat{X}_{T+(n \times 24 \times 7)}}
\end{displaymath}

 \vspace{1em}
\noindent
\includegraphics[width=.95\linewidth]{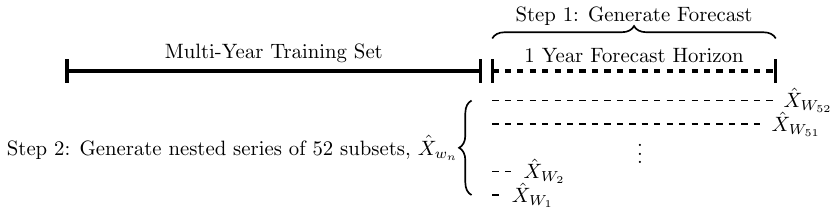}

\textbf{Step 3:} Compute the optimum commitment level $c_{w_n}$ with minimal cost over the future data set $\hat{X}_{w_n}$ for each of the forecast subsets from the previous step.

\begin{displaymath}
c_{w_n} = \operatorname*{arg\,min}_c C(c, \hat{X}_{w_n})
\end{displaymath}

\noindent
\includegraphics[width=.95\linewidth]{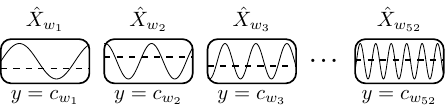}

\textbf{Step 4:} Take the minimum of each of these commitment levels as the amount to set as the commitment level for the coming week.

\begin{displaymath}
c^* = \min\{c_{w_1}, c_{w_2}, \ldots, c_{w_{52}}\}
\end{displaymath} 
\caption{Optimal Commitment For Demand Forecast}

\label{alg:forecast}
\end{algorithm}

\begin{figure}
    \centering
    \includegraphics[width=\linewidth]{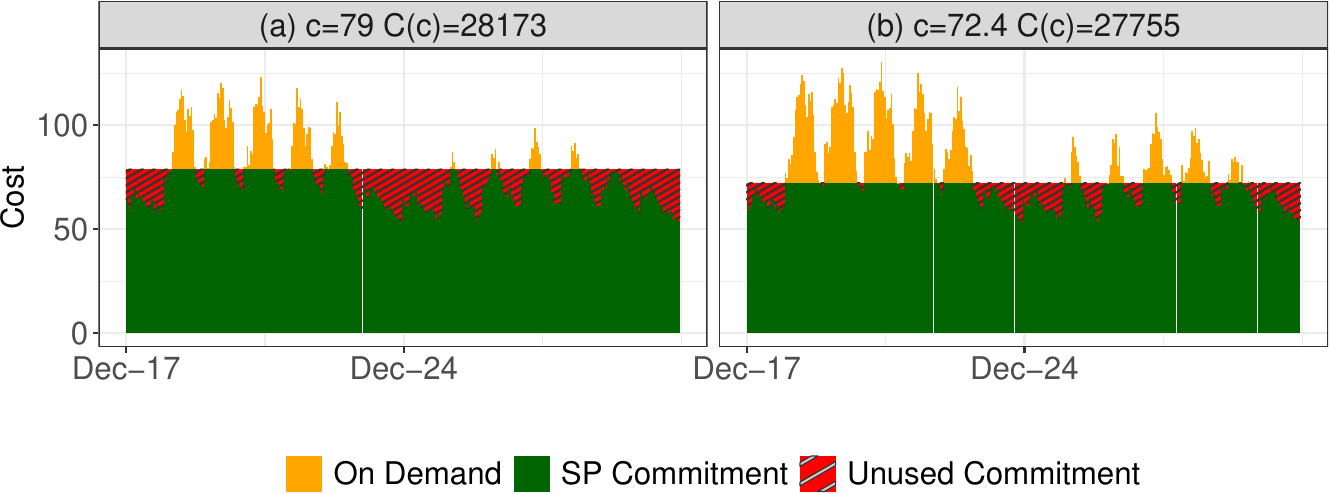}
    
    \begin{minipage}[t]{0.48\linewidth}
        \subcaption{The commitment level $c_{w_1}=79$ is the minimal cost $C(c, \hat{X}_{w_1})$ commitment level when considering 1 week into the future.  When applying this commitment across both weeks, the total cost is \$28,173.}
    \end{minipage}
    \hfill
    \begin{minipage}[t]{0.48\linewidth}
        \subcaption{The commitment level $c_{w_2}=72.4$ is the minimal cost $C(c, \hat{X}_{w_2})$ commitment level when considering 2 weeks into the future.  When applying this commitment across both weeks, the total cost is \$27,755.}
    \end{minipage}

    \caption{Comparison of $C(c_{w_1}, \hat{X}_{w_2})$ vs $C(c_{w_2}, \hat{X}_{w_2})$ for commitment-levels computed over 1-week vs 2-weeks forecast horizon to illustrate the importance of considering future reductions in demand.}
    \Description{Two side by side plots showing the VM cost data from December 17 to December 31.  The commitment level is higher for the left plot and the captions provide more details on the total costs for each scenario.}
    
    \label{fig:side-by-side}
\end{figure}

Figure~\ref{fig:side-by-side} illustrates Steps 3 and 4 for an example two-week demand forecast generated on December 16, 2023.  The left example considers only the next week forecast, $\hat{X}_{w_1}$ and identifies an optimal commitment level of 79.  The right example uses the two-week forecast, $\hat{X}_{w_2}$, to calculate a lower optimal commitment level of 72.4 because of the forecasted drop in demand during the holiday week December 24-30.  Using this two-week forecast produces a lower total cost for the forecasted compute demand over the entire 2-week time horizon.


\subsubsection{Laddering}

Bond laddering has long been used to help investors respond to changing interest rates and capital needs \cite{judd_bond_2002, laddering}, and the same advantages apply to managing the risk of long-term capacity commitments.
Purchasing commitments with staggered start and end dates provides flexibility.
This approach combines the benefits of increased discounts from longer-term commitments with the flexibility offered by staggered expiration dates.
Each incremental capacity commitment purchase represents a step-function increase in commitment and these increments can be applied regularly to match the overall increasing trend of demand.  
This upward ramp in cumulative commitment level that is generated by incremental purchases yields a rolling downward expiration at the conclusion of each laddered commitment's term.
These expirations are fundamental to proactively scaling down commitment. For example, if we expect VM demand to decrease due to a new hardware or software performance optimization, we simply stop purchasing new commitments. This pause allows existing commitments to expire.  Laddering begins as a long-term strategy, as the full benefit is realized only when coverage expires consistently.

\begin{figure}
    \centering
    \includegraphics[width=\linewidth]{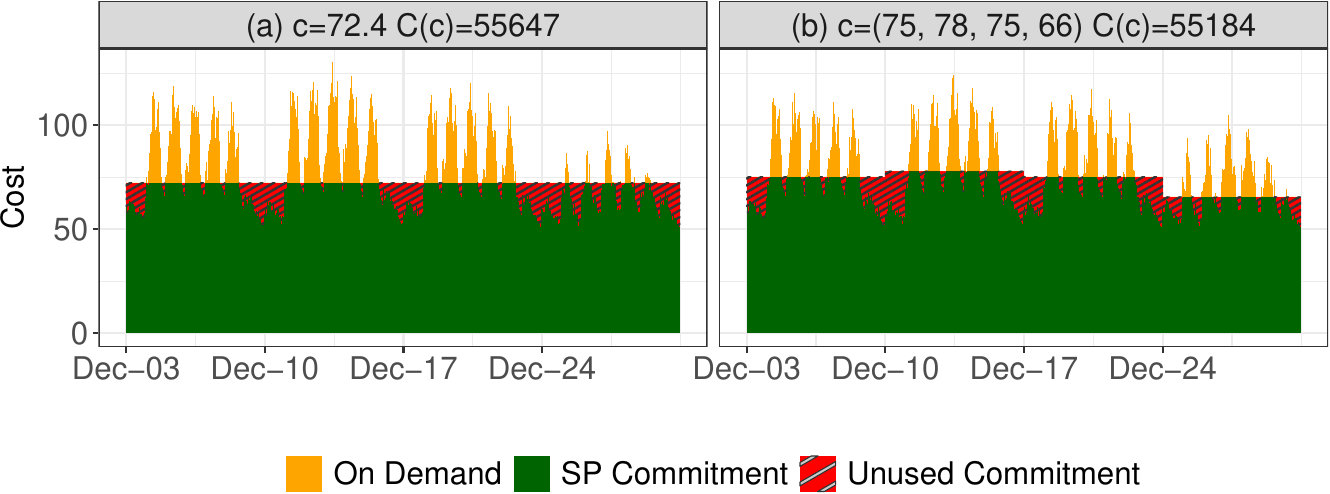}
 
    \begin{minipage}[t]{0.48\linewidth}
    \subcaption{The commitment level $c_{w_4}=72.4$ is the minimal cost $C(c, \hat{X}_{w_4})$ commitment level considering 4 weeks into the future.  When applying this commitment across all four weeks, the total cost is \$55,647.}
    \end{minipage}
    \hfill
    \begin{minipage}[t]{0.48\linewidth}
    \subcaption{In the best-case scenario where previously laddered commitments are expiring in the coming weeks and we can adjust the weekly commitment level to match the forecasted demand, the total cost reduces to \$55,184.}
    \end{minipage}

  \caption{Comparison of commitment levels over 4 weeks with the benefit of expiring commitments to allow commitment levels to reduce the optimal forecast level.}
    \label{fig:laddering}
    \Description{Two side by side plots showing the VM cost data from December 3 to December 31.  The commitment level is static for the left plot but the commitment level changes each week for the right plot representing the impact of expiring savings plans and additional commitment purchases to keep the commitment-level closer to the actual demand.  The captions provide more details on the total costs for each scenario.}
\end{figure}

Figure~\ref{fig:laddering} demonstrates the benefits of laddering during the year-end four-week period, which experiences a significant seasonal drop in capacity demand.
Scenario A shows a flat commitment level calculated by the algorithm from the previous section. Scenario B leverages expiring commitments to align coverage with demand.
Algorithm~\ref{alg:forecast} can be modified to accept a schedule of expiring commitments and translate optimal commitment levels into incremental purchase amounts for each period.
Assuming perfect laddering, where commitment levels are optimal each week, Scenario B achieves 1.1\% cost savings by reducing commitment before anticipated demand drops, eliminating the risk of over-commitment.
Ultimately, laddering aligns fixed capacity commitments with dynamic usage, protecting against demand reductions due to holidays or hardware improvements, and enabling aggressive cost-saving purchase strategies.

\section{Time Shifting}
\label{sec:timeshifting}

The workloads presented here exhibit a diurnal load pattern resulting from how users interact with Snowflake products supported by underlying services.  This behavior manifests in peaks and valleys of cloud compute demand as shown in Figure~\ref{fig:dailypattern}.  The previous sections focused on calculating an optimal capacity commitment level, somewhere in between the peaks and valleys that minimizes the total amount spent on cloud resources.

When workloads are at least partially fungible, however, there is another technique to greatly reduce costs.  We focus on reducing peak demand by moving a portion of the work from the peak to valley (e.g. off-peak time) which we refer in this paper as \emph{time shifting} \cite{parthtimeshifting}.

Based on the 3-year workload trace in the previous section, we find that the optimal capacity commitment leaves about 4.3\% of the capacity as unused savings plans where demand is below the long-term commitment level.  This supply of unused capacity is concentrated in a 48 hour period each weekend and a smaller amount of capacity available for 4 hours each weeknight.  This motivates the next question of which workloads can be timeshifted to utilize this off-peak free capacity.

Following Sukprasert, et al.\ \cite{temporalshifting} we use two main dimensions to characterize the temporal flexibility of a workload: deferrability and interruptibility.  Using these criteria, customer data warehouse queries are not good candidates for time shifting.  However, there are many other internal workloads that are both deferrable and interruptible and thus good candidates for time shifting:

\begin{enumerate}
\item \textbf{Automated Performance and Regression Tests}.  We use the Snowtrail \cite{snowtrail} framework to rerun customer queries on isolated virtual warehouses.  This allows us to extensively test new software releases and measure the performance of new features.  Many of these workloads can be time shifted to run during nights and weekends when we have excess free capacity.

\item \textbf{Load Tests of Internal Infrastructure}.  We run load tests on isolated pieces of infrastructure during off-peak hours to understand bottlenecks and improve our capacity models.  This allows us to avoid more expensive On-Demand capacity by utilizing the troughs of unused capacity commitments.

\item \textbf{Trust Center Security Essentials Scans}.  This background task checks for account configurations recommended by Snowflake, and asserts that multi-factor authentication (MFA) is turned on correctly for accounts, and other security policies are set.

\item \textbf{Automated Builds}.  We build many different types of artifacts for the major cloud environments from our CI/CD \cite{ci} system, and many of these builds can be run overnight so that new code is ready for use the next morning.

\end{enumerate}

In October 2024 we completed a survey of workloads across Snowflake and found that approximately 5\% of our total workloads are candidates for time shifting from among these categories, which aligns well with the 4.3\% of available unused long-term commitments.

\section{Free Pools}
\label{sec:freepool}

The previous sections described the overall demand drivers for Snowflake compute usage and how we optimize long-term capacity commitments for this base level of demand to reduce costs.  In this section we turn our attention to latency.  One of the benefits of Snowflake over traditional data warehouses is the separation of storage from compute. This enables our customers to instantaneously scale their compute resources on-demand based on their workloads.  

\begin{figure}
  \centering
  \includegraphics[width=\linewidth]{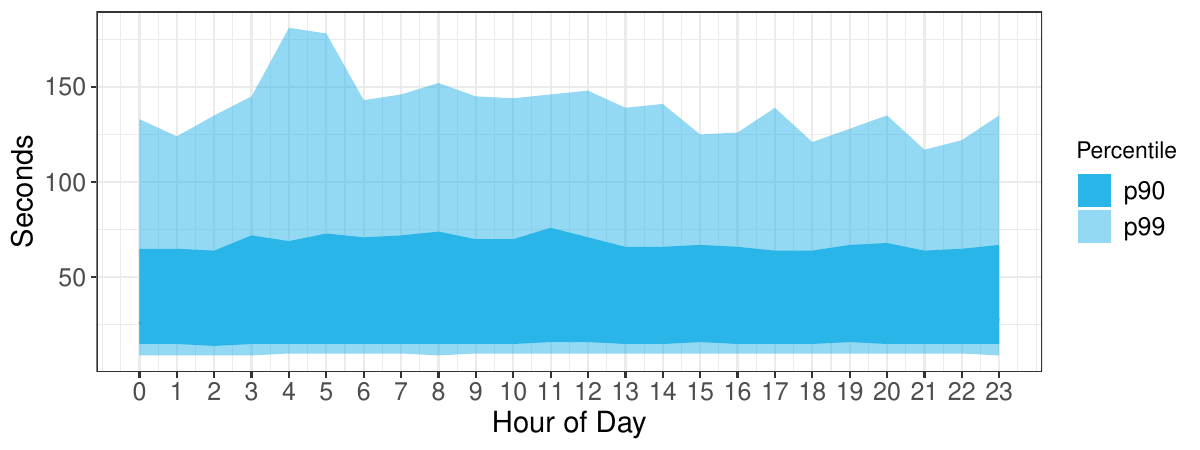}
  \caption{Observed provisioning latency by hour of day across all three cloud providers on a typical day.}
  \label{fig:vmlatency}
  \Description{A timeseries area plot over 24 hours of a day showing the p90 and p99 seconds of VM provisioning latency.  The p90 stays around 60 seconds throughout the day while the p99 stays around 120 seconds.}
\end{figure}

Snowflake uses VMs to provide the requisite multi-tenant security and resource isolation.  This provides greater security and resource isolation compared to using containers, but creates the challenge of dealing with variance in the end-to-end startup time that would be unacceptable to customers.  Figure~\ref{fig:vmlatency} shows the 90th and 99th percentile of VM provisioning latency observed throughout the day across three major clouds.  These values in the range of minutes are significantly higher than the SLO provided to Snowflake users.

When cloud providers are unable to provide resources in time to meet our Service Level Objective (SLO), we must maintain a \emph{free pool} of already provisioned capacity that is available to allow customer workloads to scale up instantaneously.
Free pools are also called ``warm pools'' \cite{jonas2019cloud}, ``live pools''
\cite{intelligentpooling}, or ``intelligent pools''  \cite{intelligentvm}. Their use is prevalent in the industry, and finding ways to efficiently manage them is an active area of research across public and private clouds. Past work describes the use of free pools for supporting the operation of Snowflake's own control plane \cite{snowflakecp}.

\subsection{Predictive Pre-Provisioning}
To insulate customers from this latency of VM provisioning, we closely track consumption and latency of VM provisioning with the cloud providers.  We combine this information with statistical forecasts of resource demand. 
Based on this information, we maintain a free pool of ready-to-go VMs that can be immediately assigned to a customer’s workloads when needed. 

Since servers in the free pools are not performing useful work for customers, they represent a cost to Snowflake. Therefore, our goal is to satisfy customer demand while minimizing the size of the free pool. This is formulated as an optimization problem to minimize free pool server-minutes while maintaining our tight internal latency targets for warehouse creation and resumption.

We model customer workloads to ensure that we maintain an optimal free pool size. Economies of scale mean that with more customer workloads on Snowflake, we are better able to absorb different types of spikes at low cost, since those spikes are usually uncorrelated. However, there are still cases where requests tend to be batched up in quick succession. For example, many customers run hourly jobs at the top of the hour or daily jobs at midnight and this can lead to a large spike of requests for more resources as can be seen over an example week in Figure 2. This problem is exacerbated by the fact that other customers of the cloud providers often have similar affinity for these time boundaries and the associated latency metrics spike at exactly the time we need them the most.

Unlike in the previous sections where we studied the total VM server-hour consumption over time, for free pools the focus is on the number of VM requested and assigned to customer workloads and of VMs being returned when workload compute is scaled down. Any analysis of per-VM usage and potential patterns therein are out of scope for this paper. 


\begin{figure}
  \centering
  \includegraphics[width=\linewidth]{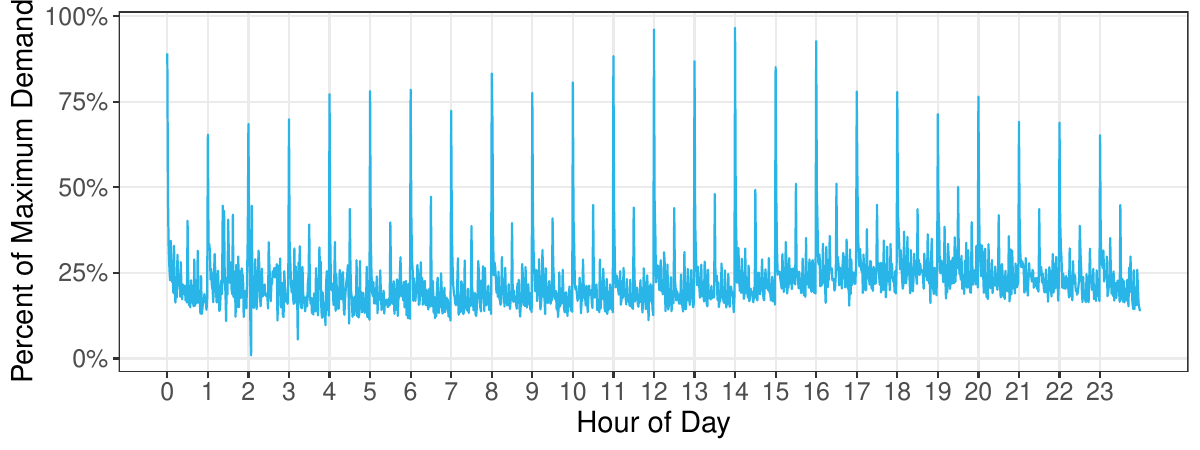}
  \caption{Example of number of VMs requested per minute for one day on a deployment.}
  \label{fig:vmcount}
  \Description{A timeseries plot over 24 hours showing the percent of maximum demand for VMs experienced throughout the day.  There are noticeable large spikes up about 60\% on the hour every hour.}
\end{figure}

Figure~\ref{fig:vmcount} shows significant temporal variation in the number of VMs requested per minute across an example day on an example deployment.  Using this historical data, we forecast the required size of the free pool with a time horizon of up to 1 hour into the future. As mentioned above, we must size our free pools to satisfy the customer demand but also be cost-efficient. So, we define our loss or cost function $c(t)$ as:

$$
c(t) = p_o \cdot \textrm{max}(0, \hat{y_t} - d_t ) + p_u \cdot \textrm{max}(0, d_t - \hat{y_t})
$$

\begin{itemize}
\item $p_o$ is the penalty we assign to each over-provisioned server
\item $p_u$ is the penalty for each under-provisioned server
\item $\hat{y}_t$ is the pool size we set
\item $d_t$ is the customer-demand of servers for a period of time t
\end{itemize}

Our goal is to forecast the value of $\hat{y}_t$ at each time period so that it minimizes our cost function $c(t)$.

Given this formulation of the cost function and the relevant inputs, we use time series forecasting to generate optimal free pool sizes, for each cloud provider deployment and for each time window. Our cloud provisioning systems then use this to request the VMs and prepare them for customer workloads before the expected customer demand requests them. By better predicting the required number of servers, our customers can scale their compute-usage in less than a second, thus dramatically speeding up their query performance. 

\begin{figure}
  \centering
  \includegraphics[width=\linewidth]{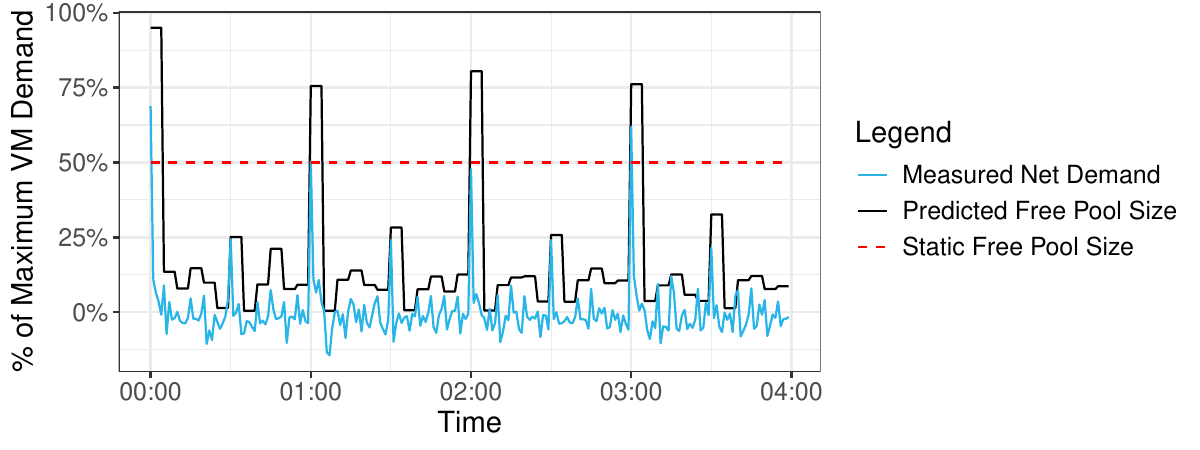}
  \caption{Comparison of static free pool and predicted free pool sizing for measured net demand of an example deployment.}
  \label{fig:freepooldone}
  \Description{A timeseries plot over four hours showing the predicted free pool size matching the measured net demand closely, including large spikes on the hour.  This compares favorably to a static free pool size which is just a single horizontal line throughout the four hour period.}
\end{figure}

The results of this prediction for an example deployment are shown in Figure~\ref{fig:freepooldone}. Over-provisioning of servers for the free pool results in wasted compute server-hours, while under-provisioning affects the customer experience and their ability to scale their compute resources. Unless the static pool size is set to be greater than the maximum VM demand, a scenario that is prohibitively costly, a static free pool size would lead to both over- and under-provisioning. Setting the static pool size at different values will only determine how much of each scenario we would encounter. For the case of predicted free pool size, apart from a performant forecaster, having the ability to rapidly modify the pool size, allows for increased cost efficiency by maintaining the exact size of required VMs at any given moment. The ability to modify the pool size is bound primarily by the provisioning latency of a given CSP and secondarily by the deprovisioning latency.

Finally, we provide a few lessons learned. First, when applying forecasting methods that take as input only a single time series, the most popular methods perform similarly. The difference in performance is mostly witnessed in how quickly or slowly they adapt to changes in the demand pattern. 
Second, there is no clear preference to a method that adapts faster or slower to the latest change in demand, since the latter can be either a new workload or a circumstantial outlier. 
Third, forecasting quality increases when introducing auxiliary data and information. 
Furthermore, when putting an ML-based feature in ``production'', the forecasting component is only one of many required building blocks. Equally, if not more, important is implementing an enclosing framework for orchestrating the free pool size management, in the presence of both external (e.g. CSP outages) and internal (e.g. extraordinary migrations) special scenarios. 
Lastly, Snowflake has grown its cloud footprint significantly over the last few years. As a result, oftentimes our system's actions to accommodate our customers' demand causes an increased load for CSPs that exposes bottlenecks or limitations in their infrastructure.

\section{Dataset}
\label{sec:dataset}

To support further research into cloud forecasting, commitment optimization, and capacity planning, we have released a data artifact of
normalized Virtual Machine (VM) demand for 12 different machine
types in 4 different regions over a 3-year period from the
Snowflake Data Cloud.

The dataset is archived on Zenodo for long-term preservation \cite{shavedicedata} and is readily available on GitHub for ease of use: 
\url{https://github.com/Snowflake-Labs/shavedice-dataset}.  The exact cloud provider, region names, and virtual machine types are obfuscated, and the total volumes are normalized.
To ensure reproducibility, code to generate most of the figures in this paper are included with the dataset.

\section{Related Work}

This section surveys related work in the field of cloud capacity planning and optimizing costs for large cloud workloads.

\subsection{Cloud Capacity Planning}

The influential position paper \cite{abovetheclouds} highlighted the connection between capacity requirements, hardware utilization, hardware generational improvements, and diurnal and seasonal workload cycles and shifting risk from SaaS Providers to Cloud Providers.  However, this paper predates the introduction of Reserved Instances, Savings Plans, and other forms of long-term commitment discounts that CSPs now offer and therefore, it does not discuss any tradeoffs of risk and cost between short and long-term commitments.  This work also doesn't provide detailed analysis of workloads or CSP performance in provisioning VMs.

More recent work such as \cite{noreservations} highlights the cost benefits of entering into multi-year capacity commitments, and analyzes the prices available for reselling long-term commitments to mitigate the risk of expected future demand not materializing.
We believe these techniques are orthogonal to the optimal purchase strategies described in this paper.
Additionally, reselling long-term commitments is not always possible due to scale, contractual limitations, or specialized resource needs without general demand from other potential customers.

Our work is closest to \cite{toreserve} in that we seek to optimize the amount and timing of long-term capacity commitments to reduce costs.  However, unlike this work, we find that the Snowflake workloads have clear daily, weekly, and seasonal periodicity over a multi-year timeframe, as well as predictable increases in hardware and software efficiency making it crucial to accurately reflect this demand.  Commercial tools such as \cite{prosperops,missioncloud} similarly suffer from an inability to automatically incorporate the known signals of upcoming hardware transitions, software rearchitecture and performance work, and new feature launches to make optimal forward-looking capacity commitments.


\subsection{Time Shifting}

There has been recent work on time shifting workloads, particularly for the purpose of reducing carbon emissions.  Much of this research characterizes the regional and temporal variation of carbon-intensity of energy supplies and studies methods of load-shifting that demand to reduce overall emissions.  There has been a dearth of realistic studies from the perspective of CSP consumers because openly available cloud computing data sets do not contain information about the delay-tolerance of the workloads.

Some representative techniques for exploiting temporal flexibility are delaying workload execution \cite{cirnetimeshifting}, using suspend-resume techniques \cite{letswait}, or scaling the workload \cite{carbonscaler}. 
These three papers focus on matching workloads to the supply of green energy in large commercial data centers.
The first paper demonstrates a 1-2\% power reduction for aggregate Google data center workloads by setting constraints during peak carbon intensity periods.
The second characterizes carbon-intensity using simulated and batch ML training data.
The final paper relies on a greedy autoscaling algorithm minimizing emissions via resource scaling, evaluated on ML training traces.

None of these papers characterizes the workload temporal patterns with the level of detail of \S~\ref{sec:userdemand}, and most rely on simulations or very fungible batch workloads.
Studies by \cite{parthtimeshifting} and \cite{brookstimeshifting} explore time shifting of workloads in Facebook data centers.  In the latter study, the authors find that CPU utilization swings by about 20\% throughout the day in Facebook data centers, but power consumption only varies by 4\%, on average.  This highlights the stark difference between the relatively large and immediate impact of time shifting demand from peak to trough for a cloud customer with long-term capacity commitments, compared to the relatively smaller impact of time shifting demand in physical data centers to reduce electricity consumption.


\subsection{Predictive Pre-Provisioning}

The concept of preparing in advance a set of units for computation, in order to reduce the latency to fetch them, is prevalent across computer science, with memory/data caching being one of the most obvious applications. For cloud computing, the majority of work has focused on VM or container placement in the cloud-providers' datacenters \cite{khazaei2012analysis,tseng2017dynamic}, or pre-warming of machines for lower latency execution of workloads on serverless platforms \cite{brooker2023demand,li2022help}. 
In \cite{mlvmprovisioning,intelligentpooling} the authors describe using ML to pre-provision VMs to offer improved SLAs for cloud workloads.  \cite{workloadpredictionarima} builds a comprehensive workload prediction ARIMA model and uses it to dynamically provision VMs in the cloud.
Overall, most of the existing work has focused on the optimization of the cloud providers' infrastructure and QoS; we refer to \cite{gao2020machine} for a comprehensive list of existing ML-related work focusing on CSPs. To the best of our knowledge, there are no other broad studies showing the real-world performance of VM provisioning from cloud providers.




\section{Conclusions}
This paper characterizes aggregate compute demand patterns across a large multi-cloud platform encompassing data warehousing, data lakes, data science, data engineering, and machine learning workloads.
First, we analyzed a three year trace of compute demand and identified daily, weekly, and seasonal patterns in demand.  We also examined multiple years of price and performance data from three major CSPs and the software performance improvements made to Snowflake.  Using this, we analyzed the impact of these trends on long-term capacity commitments and formulated an optimization problem to identify the optimal level of long-term capacity commitment.  We then characterized a number of internal workloads at Snowflake that are amenable to time shifting to further reduce costs.  We also analyzed the performance of the CSPs at providing VMs at large scale and formulated a second optimization problem to minimize costs for maintaining a pool of pre-provisioned VMs that improves performance. 
To support further research in this area, we have released a public dataset of normalized cloud capacity demand from the Snowflake Data Cloud.
The techniques described here have been implemented within Snowflake, reducing cloud computing costs.

\section{Future Work}

The workload analysis and optimization formulations presented in this paper establish a robust framework for balancing cost savings and demand risk in large-scale, multi-cloud compute capacity commitments.
While incremental refinements to our existing demand optimizations are possible, we believe the most significant gains will stem from actively shaping demand patterns through \emph{time shifting strategies}.
Reducing the inherent "spikiness" of workloads can substantially enhance efficiency and cost-effectiveness.  Consequently, more research is needed in the following key areas:

\begin{enumerate}
\item \textbf{Capacity-Commitment-Aware Scheduling of Deferrable Workloads:}  Effectively leveraging capacity commitments necessitates sophisticated workload scheduling that can predict and dynamically adapt to fluctuating demand.  A key challenge lies in accurately characterizing and predicting the deferability of workloads at large scale.  While algorithms exist for basic time shifting, the industry lacks solutions for identifying dependencies and the true operational impact of delaying specific tasks.  This includes understanding the trade-offs between workload priority, service level agreements, and the potential cost savings opportunities.
\item \textbf{Economic Incentives:} In~\cite{resourcemarkets} one of the authors explored using market-pricing incentives to migrate internal workloads from congested regions to under-utilized regions.  We plan to explore these ideas in the context of time shifting to provide incentives for internal engineering teams to migrate workloads from peak time periods where we are paying on-demand rates, to off-peak times on evenings and weekends when our total demand is below the savings plan commitment levels for the major cloud providers.
\end{enumerate}

\begin{acks}
This work would not have been possible without the help of many individuals and teams inside Snowflake.
We are especially grateful to Eric Nguyen, Karan Vishwanathan, Angel Xiong, Bradley Jiang, Mustafa Bal, and Xiaojun Zhao for their help applying these ideas in production.
Gabe Bryant and Mike Meyer also contributed ideas and improvements to the presentation.
\end{acks}

\bibliographystyle{ACM-Reference-Format}
\bibliography{refs}


\begin{thebibliography}{54}


\ifx \showCODEN    \undefined \def \showCODEN     #1{\unskip}     \fi
\ifx \showDOI      \undefined \def \showDOI       #1{#1}\fi
\ifx \showISBNx    \undefined \def \showISBNx     #1{\unskip}     \fi
\ifx \showISBNxiii \undefined \def \showISBNxiii  #1{\unskip}     \fi
\ifx \showISSN     \undefined \def \showISSN      #1{\unskip}     \fi
\ifx \showLCCN     \undefined \def \showLCCN      #1{\unskip}     \fi
\ifx \shownote     \undefined \def \shownote      #1{#1}          \fi
\ifx \showarticletitle \undefined \def \showarticletitle #1{#1}   \fi
\ifx \showURL      \undefined \def \showURL       {\relax}        \fi
\providecommand\bibfield[2]{#2}
\providecommand\bibinfo[2]{#2}
\providecommand\natexlab[1]{#1}
\providecommand\showeprint[2][]{arXiv:#2}

\bibitem[mis(2024)]%
        {missioncloud}
 \bibinfo{year}{2024}\natexlab{}.
\newblock \bibinfo{booktitle}{\emph{Mission Cloud}}.
\newblock
\urldef\tempurl%
\url{https://www.missioncloud.com}
\showURL{%
\tempurl}


\bibitem[pro(2024)]%
        {prosperops}
 \bibinfo{year}{2024}\natexlab{}.
\newblock \bibinfo{booktitle}{\emph{ProsperOps}}.
\newblock
\urldef\tempurl%
\url{https://www.prosperops.com}
\showURL{%
\tempurl}


\bibitem[Acun et~al\mbox{.}(2023)]%
        {brookstimeshifting}
\bibfield{author}{\bibinfo{person}{Bilge Acun}, \bibinfo{person}{Benjamin Lee},
  \bibinfo{person}{Fiodar Kazhamiaka}, \bibinfo{person}{Kiwan Maeng},
  \bibinfo{person}{Udit Gupta}, \bibinfo{person}{Manoj Chakkaravarthy},
  \bibinfo{person}{David Brooks}, {and} \bibinfo{person}{Carole-Jean Wu}.}
  \bibinfo{year}{2023}\natexlab{}.
\newblock \showarticletitle{Carbon Explorer: A Holistic Framework for Designing
  Carbon Aware Datacenters}. In \bibinfo{booktitle}{\emph{Proceedings of the
  28th ACM International Conference on Architectural Support for Programming
  Languages and Operating Systems, Volume 2}} (Vancouver, BC, Canada)
  \emph{(\bibinfo{series}{ASPLOS 2023})}. \bibinfo{publisher}{Association for
  Computing Machinery}, \bibinfo{address}{New York, NY, USA},
  \bibinfo{pages}{118–132}.
\newblock
\showISBNx{9781450399166}
\urldef\tempurl%
\url{https://doi.org/10.1145/3575693.3575754}
\showDOI{\tempurl}


\bibitem[Amazon(2024a)]%
        {awsprice}
\bibfield{author}{\bibinfo{person}{Amazon}.} \bibinfo{year}{2024}\natexlab{a}.
\newblock \bibinfo{booktitle}{\emph{{Amazon EC2 On-Demand Pricing (US East
  Ohio)}}}.
\newblock
\urldef\tempurl%
\url{https://aws.amazon.com/ec2/pricing/on-demand/}
\showURL{%
Retrieved October 24, 2024 from \tempurl}


\bibitem[Amazon(2024b)]%
        {aws}
\bibfield{author}{\bibinfo{person}{Amazon}.} \bibinfo{year}{2024}\natexlab{b}.
\newblock \bibinfo{booktitle}{\emph{Amazon Web Services}}.
\newblock
\urldef\tempurl%
\url{https://aws.amazon.com/}
\showURL{%
\tempurl}


\bibitem[Amazon(2024c)]%
        {awsg3perf}
\bibfield{author}{\bibinfo{person}{Amazon}.} \bibinfo{year}{2024}\natexlab{c}.
\newblock \bibinfo{booktitle}{\emph{{AWS Unveils Next Generation AWS-Designed
  Chips}}}.
\newblock
\urldef\tempurl%
\url{https://aws.amazon.com/blogs/aws/new-graviton3-based-general-purpose-m7g-and-memory-optimized-r7g-amazon-ec2-instances/}
\showURL{%
Retrieved October 28, 2024 from \tempurl}


\bibitem[Amazon(2024d)]%
        {awsg4perf}
\bibfield{author}{\bibinfo{person}{Amazon}.} \bibinfo{year}{2024}\natexlab{d}.
\newblock \bibinfo{booktitle}{\emph{{AWS Unveils Next Generation AWS-Designed
  Chips}}}.
\newblock
\urldef\tempurl%
\url{https://press.aboutamazon.com/2023/11/aws-unveils-next-generation-aws-designed-chips}
\showURL{%
Retrieved October 28, 2024 from \tempurl}


\bibitem[Amazon(2024e)]%
        {awsg3perf1}
\bibfield{author}{\bibinfo{person}{Amazon}.} \bibinfo{year}{2024}\natexlab{e}.
\newblock \bibinfo{booktitle}{\emph{{Powering Amazon RDS with AWS Graviton3:
  Benchmarks}}}.
\newblock
\urldef\tempurl%
\url{https://aws.amazon.com/blogs/database/powering-amazon-rds-with-aws-graviton3-benchmarks/}
\showURL{%
Retrieved October 31, 2024 from \tempurl}


\bibitem[Ambati et~al\mbox{.}(2020)]%
        {noreservations}
\bibfield{author}{\bibinfo{person}{Pradeep Ambati}, \bibinfo{person}{David
  Irwin}, {and} \bibinfo{person}{Prashant Shenoy}.}
  \bibinfo{year}{2020}\natexlab{}.
\newblock \showarticletitle{No Reservations: A First Look at
  Amazon{\textquoteright}s Reserved Instance Marketplace}. In
  \bibinfo{booktitle}{\emph{12th USENIX Workshop on Hot Topics in Cloud
  Computing (HotCloud 20)}}. \bibinfo{publisher}{USENIX Association}.
\newblock


\bibitem[Armbrust et~al\mbox{.}(2010)]%
        {abovetheclouds}
\bibfield{author}{\bibinfo{person}{Michael Armbrust}, \bibinfo{person}{Armando
  Fox}, \bibinfo{person}{Rean Griffith}, \bibinfo{person}{Anthony~D. Joseph},
  \bibinfo{person}{Randy Katz}, \bibinfo{person}{Andy Konwinski},
  \bibinfo{person}{Gunho Lee}, \bibinfo{person}{David Patterson},
  \bibinfo{person}{Ariel Rabkin}, \bibinfo{person}{Ion Stoica}, {and}
  \bibinfo{person}{Matei Zaharia}.} \bibinfo{year}{2010}\natexlab{}.
\newblock \showarticletitle{A View of Cloud Computing}.
\newblock \bibinfo{journal}{\emph{Commun. ACM}} \bibinfo{volume}{53},
  \bibinfo{number}{4} (\bibinfo{date}{April} \bibinfo{year}{2010}),
  \bibinfo{pages}{50–58}.
\newblock
\showISSN{0001-0782}
\urldef\tempurl%
\url{https://doi.org/10.1145/1721654.1721672}
\showDOI{\tempurl}


\bibitem[Barroso(2005)]%
        {priceperformance}
\bibfield{author}{\bibinfo{person}{Luiz~Andr\'{e} Barroso}.}
  \bibinfo{year}{2005}\natexlab{}.
\newblock \showarticletitle{The Price of Performance: An Economic Case for Chip
  Multiprocessing}.
\newblock \bibinfo{journal}{\emph{Queue}} \bibinfo{volume}{3},
  \bibinfo{number}{7} (\bibinfo{date}{Sept.} \bibinfo{year}{2005}),
  \bibinfo{pages}{48–53}.
\newblock
\showISSN{1542-7730}
\urldef\tempurl%
\url{https://doi.org/10.1145/1095408.1095420}
\showDOI{\tempurl}


\bibitem[Brent(1973)]%
        {brent1973algorithms}
\bibfield{author}{\bibinfo{person}{Richard~P. Brent}.}
  \bibinfo{year}{1973}\natexlab{}.
\newblock \bibinfo{booktitle}{\emph{Algorithms for Minimization without
  Derivatives}}.
\newblock \bibinfo{publisher}{Prentice-Hall}, \bibinfo{address}{Englewood
  Cliffs, NJ}.
\newblock


\bibitem[Brooker et~al\mbox{.}(2023)]%
        {brooker2023demand}
\bibfield{author}{\bibinfo{person}{Marc Brooker}, \bibinfo{person}{Mike
  Danilov}, \bibinfo{person}{Chris Greenwood}, {and} \bibinfo{person}{Phil
  Piwonka}.} \bibinfo{year}{2023}\natexlab{}.
\newblock \showarticletitle{On-demand Container Loading in $\{$AWS$\}$ Lambda}.
  In \bibinfo{booktitle}{\emph{2023 USENIX Annual Technical Conference (USENIX
  ATC 23)}}. \bibinfo{pages}{315--328}.
\newblock


\bibitem[Calheiros et~al\mbox{.}(2015)]%
        {workloadpredictionarima}
\bibfield{author}{\bibinfo{person}{Rodrigo~N. Calheiros},
  \bibinfo{person}{Enayat Masoumi}, \bibinfo{person}{Rajiv Ranjan}, {and}
  \bibinfo{person}{Rajkumar Buyya}.} \bibinfo{year}{2015}\natexlab{}.
\newblock \showarticletitle{Workload Prediction Using ARIMA Model and Its
  Impact on Cloud Applications’ QoS}.
\newblock \bibinfo{journal}{\emph{IEEE Transactions on Cloud Computing}}
  \bibinfo{volume}{3}, \bibinfo{number}{4} (\bibinfo{year}{2015}),
  \bibinfo{pages}{449--458}.
\newblock
\urldef\tempurl%
\url{https://doi.org/10.1109/TCC.2014.2350475}
\showDOI{\tempurl}


\bibitem[Carlin(2023)]%
        {prosperopscyclical}
\bibfield{author}{\bibinfo{person}{Erik Carlin}.}
  \bibinfo{year}{2023}\natexlab{}.
\newblock \bibinfo{booktitle}{\emph{Introducing Advanced Cyclical Optimization:
  Automatically Increase Compute Savings on Cyclical Cloud Workloads}}.
\newblock
\urldef\tempurl%
\url{https://www.prosperops.com/blog/introducing-advanced-cyclical-optimization-automatically-increase-compute-savings-on-cyclical-cloud-workloads/}
\showURL{%
Retrieved September 26, 2024 from \tempurl}


\bibitem[Chen et~al\mbox{.}(2022)]%
        {codeefficiency}
\bibfield{author}{\bibinfo{person}{Binghong Chen}, \bibinfo{person}{Daniel
  Tarlow}, \bibinfo{person}{Kevin Swersky}, \bibinfo{person}{Martin Maas},
  \bibinfo{person}{Pablo Heiber}, \bibinfo{person}{Ashish Naik},
  \bibinfo{person}{Milad Hashemi}, {and} \bibinfo{person}{Parthasarathy
  Ranganathan}.} \bibinfo{year}{2022}\natexlab{}.
\newblock \bibinfo{title}{Learning to Improve Code Efficiency}.
\newblock
\newblock
\showeprint[arxiv]{2208.05297}~[cs.SE]
\urldef\tempurl%
\url{https://arxiv.org/abs/2208.05297}
\showURL{%
\tempurl}


\bibitem[Dageville et~al\mbox{.}(2016)]%
        {snowflake}
\bibfield{author}{\bibinfo{person}{Benoit Dageville}, \bibinfo{person}{Thierry
  Cruanes}, \bibinfo{person}{Marcin Zukowski}, \bibinfo{person}{Vadim Antonov},
  \bibinfo{person}{Artin Avanes}, \bibinfo{person}{Jon Bock},
  \bibinfo{person}{Jonathan Claybaugh}, \bibinfo{person}{Daniel Engovatov},
  \bibinfo{person}{Martin Hentschel}, \bibinfo{person}{Jiansheng Huang},
  \bibinfo{person}{Allison~W. Lee}, \bibinfo{person}{Ashish Motivala},
  \bibinfo{person}{Abdul~Q. Munir}, \bibinfo{person}{Steven Pelley},
  \bibinfo{person}{Peter Povinec}, \bibinfo{person}{Greg Rahn},
  \bibinfo{person}{Spyridon Triantafyllis}, {and} \bibinfo{person}{Philipp
  Unterbrunner}.} \bibinfo{year}{2016}\natexlab{}.
\newblock \showarticletitle{The Snowflake Elastic Data Warehouse}. In
  \bibinfo{booktitle}{\emph{Proceedings of the 2016 International Conference on
  Management of Data}} (San Francisco, California, USA)
  \emph{(\bibinfo{series}{SIGMOD '16})}. \bibinfo{publisher}{Association for
  Computing Machinery}, \bibinfo{address}{New York, NY, USA},
  \bibinfo{pages}{215–226}.
\newblock
\showISBNx{9781450335317}
\urldef\tempurl%
\url{https://doi.org/10.1145/2882903.2903741}
\showDOI{\tempurl}


\bibitem[Gao et~al\mbox{.}(2020)]%
        {gao2020machine}
\bibfield{author}{\bibinfo{person}{Jiechao Gao}, \bibinfo{person}{Haoyu Wang},
  {and} \bibinfo{person}{Haiying Shen}.} \bibinfo{year}{2020}\natexlab{}.
\newblock \showarticletitle{Machine Learning Based Workload Prediction in Cloud
  Computing}. In \bibinfo{booktitle}{\emph{2020 29th International Conference
  on Computer Communications and Networks (ICCCN)}}. IEEE,
  \bibinfo{pages}{1--9}.
\newblock


\bibitem[Gilgur et~al\mbox{.}(2021)]%
        {parthtimeshifting}
\bibfield{author}{\bibinfo{person}{Alexander Gilgur}, \bibinfo{person}{Brian
  Coutinho}, \bibinfo{person}{Iyswarya Narayanan}, {and} \bibinfo{person}{Parth
  Malani}.} \bibinfo{year}{2021}\natexlab{}.
\newblock \showarticletitle{Transitive Power Modeling for Improving Resource
  Efficiency in a Hyperscale Datacenter}. In
  \bibinfo{booktitle}{\emph{Companion Proceedings of the Web Conference 2021}}
  (Ljubljana, Slovenia) \emph{(\bibinfo{series}{WWW '21})}.
  \bibinfo{publisher}{Association for Computing Machinery},
  \bibinfo{address}{New York, NY, USA}, \bibinfo{pages}{182–191}.
\newblock
\showISBNx{9781450383134}
\urldef\tempurl%
\url{https://doi.org/10.1145/3442442.3452057}
\showDOI{\tempurl}


\bibitem[Google(2024a)]%
        {gcp}
\bibfield{author}{\bibinfo{person}{Google}.} \bibinfo{year}{2024}\natexlab{a}.
\newblock \bibinfo{booktitle}{\emph{Google Cloud Platform}}.
\newblock
\urldef\tempurl%
\url{https://cloud.google.com/}
\showURL{%
\tempurl}


\bibitem[Google(2024b)]%
        {gcpperf}
\bibfield{author}{\bibinfo{person}{Google}.} \bibinfo{year}{2024}\natexlab{b}.
\newblock \bibinfo{booktitle}{\emph{{Introducing Google Axion Processors, our
  new Arm-based CPUs}}}.
\newblock
\urldef\tempurl%
\url{https://cloud.google.com/blog/products/compute/introducing-googles-new-arm-based-cpu/}
\showURL{%
Retrieved October 26, 2024 from \tempurl}


\bibitem[Google(2024c)]%
        {gcpprice}
\bibfield{author}{\bibinfo{person}{Google}.} \bibinfo{year}{2024}\natexlab{c}.
\newblock \bibinfo{booktitle}{\emph{{VM instance pricing (Iowa US-Central1)}}}.
\newblock
\urldef\tempurl%
\url{https://cloud.google.com/compute/vm-instance-pricing#n2_predefined}
\showURL{%
Retrieved October 26, 2024 from \tempurl}


\bibitem[Hanafy et~al\mbox{.}(2024)]%
        {carbonscaler}
\bibfield{author}{\bibinfo{person}{Walid~A. Hanafy}, \bibinfo{person}{Qianlin
  Liang}, \bibinfo{person}{Noman Bashir}, \bibinfo{person}{David Irwin}, {and}
  \bibinfo{person}{Prashant Shenoy}.} \bibinfo{year}{2024}\natexlab{}.
\newblock \showarticletitle{CarbonScaler: Leveraging Cloud Workload Elasticity
  for Optimizing Carbon-Efficiency}.
\newblock \bibinfo{journal}{\emph{SIGMETRICS Perform. Eval. Rev.}}
  \bibinfo{volume}{52}, \bibinfo{number}{1} (\bibinfo{date}{June}
  \bibinfo{year}{2024}), \bibinfo{pages}{49–50}.
\newblock
\showISSN{0163-5999}
\urldef\tempurl%
\url{https://doi.org/10.1145/3673660.3655048}
\showDOI{\tempurl}


\bibitem[Hilton et~al\mbox{.}(2016)]%
        {ci}
\bibfield{author}{\bibinfo{person}{Michael Hilton}, \bibinfo{person}{Timothy
  Tunnell}, \bibinfo{person}{Kai Huang}, \bibinfo{person}{Darko Marinov}, {and}
  \bibinfo{person}{Danny Dig}.} \bibinfo{year}{2016}\natexlab{}.
\newblock \showarticletitle{Usage, Costs, and Benefits of Continuous
  Integration in Open-Source Projects}. In
  \bibinfo{booktitle}{\emph{Proceedings of the 31st IEEE/ACM International
  Conference on Automated Software Engineering}} (Singapore, Singapore)
  \emph{(\bibinfo{series}{ASE '16})}. \bibinfo{publisher}{Association for
  Computing Machinery}, \bibinfo{address}{New York, NY, USA},
  \bibinfo{pages}{426–437}.
\newblock
\showISBNx{9781450338455}
\urldef\tempurl%
\url{https://doi.org/10.1145/2970276.2970358}
\showDOI{\tempurl}


\bibitem[Jonas et~al\mbox{.}(2019)]%
        {jonas2019cloud}
\bibfield{author}{\bibinfo{person}{Eric Jonas}, \bibinfo{person}{Johann
  Schleier-Smith}, \bibinfo{person}{Vikram Sreekanti},
  \bibinfo{person}{Chia-Che Tsai}, \bibinfo{person}{Anurag Khandelwal},
  \bibinfo{person}{Qifan Pu}, \bibinfo{person}{Vaishaal Shankar},
  \bibinfo{person}{Joao Carreira}, \bibinfo{person}{Karl Krauth},
  \bibinfo{person}{Neeraja Yadwadkar}, {et~al\mbox{.}}}
  \bibinfo{year}{2019}\natexlab{}.
\newblock \showarticletitle{Cloud Programming Simplified: A Berkeley View on
  Serverless Computing}.
\newblock \bibinfo{journal}{\emph{arXiv preprint arXiv:1902.03383}}
  (\bibinfo{year}{2019}).
\newblock


\bibitem[Judd(2002)]%
        {judd_bond_2002}
\bibfield{author}{\bibinfo{person}{Kenneth~L. Judd}.}
  \bibinfo{year}{2002}\natexlab{}.
\newblock \showarticletitle{Bond Ladders and Optimal Portfolios}.
\newblock \bibinfo{journal}{\emph{Journal of Economic Dynamics and Control}}
  \bibinfo{volume}{26}, \bibinfo{number}{9-10} (\bibinfo{year}{2002}),
  \bibinfo{pages}{1485--1514}.
\newblock
\urldef\tempurl%
\url{https://doi.org/10.1016/S0165-1889(01)00078-X}
\showDOI{\tempurl}


\bibitem[Khazaei et~al\mbox{.}(2012)]%
        {khazaei2012analysis}
\bibfield{author}{\bibinfo{person}{Hamzeh Khazaei}, \bibinfo{person}{Jelena
  Mi{\v{s}}i{\'c}}, \bibinfo{person}{Vojislav~B Mi{\v{s}}i{\'c}}, {and}
  \bibinfo{person}{Saeed Rashwand}.} \bibinfo{year}{2012}\natexlab{}.
\newblock \showarticletitle{Analysis of a Pool Management Scheme for Cloud
  Computing Centers}.
\newblock \bibinfo{journal}{\emph{IEEE Transactions on Parallel and Distributed
  Systems}} \bibinfo{volume}{24}, \bibinfo{number}{5} (\bibinfo{year}{2012}),
  \bibinfo{pages}{849--861}.
\newblock


\bibitem[Li et~al\mbox{.}(2022)]%
        {li2022help}
\bibfield{author}{\bibinfo{person}{Zijun Li}, \bibinfo{person}{Linsong Guo},
  \bibinfo{person}{Quan Chen}, \bibinfo{person}{Jiagan Cheng},
  \bibinfo{person}{Chuhao Xu}, \bibinfo{person}{Deze Zeng},
  \bibinfo{person}{Zhuo Song}, \bibinfo{person}{Tao Ma}, \bibinfo{person}{Yong
  Yang}, \bibinfo{person}{Chao Li}, {et~al\mbox{.}}}
  \bibinfo{year}{2022}\natexlab{}.
\newblock \showarticletitle{Help Rather Than Recycle: Alleviating Cold Startup
  in Serverless Computing Through $\{$Inter-Function$\}$ Container Sharing}. In
  \bibinfo{booktitle}{\emph{2022 USENIX Annual Technical Conference (USENIX ATC
  22)}}. \bibinfo{pages}{69--84}.
\newblock


\bibitem[Luo et~al\mbox{.}(2021)]%
        {intelligentvm}
\bibfield{author}{\bibinfo{person}{Chuan Luo}, \bibinfo{person}{Bo Qiao},
  \bibinfo{person}{Xin Chen}, \bibinfo{person}{Pu Zhao},
  \bibinfo{person}{Randolph Yao}, \bibinfo{person}{Hongyu Zhang},
  \bibinfo{person}{Wei Wu}, \bibinfo{person}{Andrew Zhou}, {and}
  \bibinfo{person}{Qingwei Lin}.} \bibinfo{year}{2021}\natexlab{}.
\newblock \showarticletitle{Intelligent Virtual Machine Provisioning in Cloud
  Computing}. In \bibinfo{booktitle}{\emph{Proceedings of the Twenty-Ninth
  International Joint Conference on Artificial Intelligence}} (Yokohama,
  Yokohama, Japan) \emph{(\bibinfo{series}{IJCAI'20})}. Article
  \bibinfo{articleno}{208}, \bibinfo{numpages}{8}~pages.
\newblock
\showISBNx{9780999241165}


\bibitem[Melissaris et~al\mbox{.}(2022)]%
        {snowflakecp}
\bibfield{author}{\bibinfo{person}{Themis Melissaris}, \bibinfo{person}{Kunal
  Nabar}, \bibinfo{person}{Rares Radut}, \bibinfo{person}{Samir Rehmtulla},
  \bibinfo{person}{Arthur Shi}, \bibinfo{person}{Samartha Chandrashekar}, {and}
  \bibinfo{person}{Ioannis Papapanagiotou}.} \bibinfo{year}{2022}\natexlab{}.
\newblock \showarticletitle{Elastic Cloud Services: Scaling Snowflake's Control
  Plane}. In \bibinfo{booktitle}{\emph{Proceedings of the 13th Symposium on
  Cloud Computing}} (San Francisco, California) \emph{(\bibinfo{series}{SoCC
  '22})}. \bibinfo{publisher}{Association for Computing Machinery},
  \bibinfo{address}{New York, NY, USA}, \bibinfo{pages}{142–157}.
\newblock
\showISBNx{9781450394147}
\urldef\tempurl%
\url{https://doi.org/10.1145/3542929.3563483}
\showDOI{\tempurl}


\bibitem[Microsoft(2024a)]%
        {azure}
\bibfield{author}{\bibinfo{person}{Microsoft}.}
  \bibinfo{year}{2024}\natexlab{a}.
\newblock \bibinfo{booktitle}{\emph{Azure}}.
\newblock
\urldef\tempurl%
\url{https://azure.microsoft.com/}
\showURL{%
\tempurl}


\bibitem[Microsoft(2024b)]%
        {azureprice}
\bibfield{author}{\bibinfo{person}{Microsoft}.}
  \bibinfo{year}{2024}\natexlab{b}.
\newblock \bibinfo{booktitle}{\emph{{Azure Pricing Calculator (East US)}}}.
\newblock
\urldef\tempurl%
\url{https://azure.microsoft.com/en-us/pricing/calculator/}
\showURL{%
Retrieved October 24, 2024 from \tempurl}


\bibitem[Nambiar and Poess(2006)]%
        {tpcds}
\bibfield{author}{\bibinfo{person}{Raghunath~Othayoth Nambiar} {and}
  \bibinfo{person}{Meikel Poess}.} \bibinfo{year}{2006}\natexlab{}.
\newblock \showarticletitle{The Making of TPC-DS.}
\newblock


\bibitem[Radovanović et~al\mbox{.}(2023)]%
        {cirnetimeshifting}
\bibfield{author}{\bibinfo{person}{Ana Radovanović}, \bibinfo{person}{Ross
  Koningstein}, \bibinfo{person}{Ian Schneider}, \bibinfo{person}{Bokan Chen},
  \bibinfo{person}{Alexandre Duarte}, \bibinfo{person}{Binz Roy},
  \bibinfo{person}{Diyue Xiao}, \bibinfo{person}{Maya Haridasan},
  \bibinfo{person}{Patrick Hung}, \bibinfo{person}{Nick Care},
  \bibinfo{person}{Saurav Talukdar}, \bibinfo{person}{Eric Mullen},
  \bibinfo{person}{Kendal Smith}, \bibinfo{person}{MariEllen Cottman}, {and}
  \bibinfo{person}{Walfredo Cirne}.} \bibinfo{year}{2023}\natexlab{}.
\newblock \showarticletitle{Carbon-Aware Computing for Datacenters}.
\newblock \bibinfo{journal}{\emph{IEEE Transactions on Power Systems}}
  \bibinfo{volume}{38}, \bibinfo{number}{2} (\bibinfo{year}{2023}),
  \bibinfo{pages}{1270--1280}.
\newblock
\urldef\tempurl%
\url{https://doi.org/10.1109/TPWRS.2022.3173250}
\showDOI{\tempurl}


\bibitem[Ravikumar et~al\mbox{.}(2024)]%
        {intelligentpooling}
\bibfield{author}{\bibinfo{person}{Deepak Ravikumar}, \bibinfo{person}{Alex
  Yeo}, \bibinfo{person}{Yiwen Zhu}, \bibinfo{person}{Aditya Lakra},
  \bibinfo{person}{Harsha Nagulapalli}, \bibinfo{person}{Santhosh Ravindran},
  \bibinfo{person}{Steve Suh}, \bibinfo{person}{Niharika Dutta},
  \bibinfo{person}{Andrew Fogarty}, \bibinfo{person}{Yoonjae Park},
  \bibinfo{person}{Sumeet Khushalani}, \bibinfo{person}{Arijit Tarafdar},
  \bibinfo{person}{Kunal Parekh}, {and} \bibinfo{person}{Subru Krishnan}.}
  \bibinfo{year}{2024}\natexlab{}.
\newblock \showarticletitle{Intelligent Pooling: Proactive Resource
  Provisioning in Large-scale Cloud Service}.
\newblock \bibinfo{journal}{\emph{Proc. VLDB Endow.}} \bibinfo{volume}{17},
  \bibinfo{number}{7} (\bibinfo{date}{May} \bibinfo{year}{2024}),
  \bibinfo{pages}{1618–1627}.
\newblock
\showISSN{2150-8097}
\urldef\tempurl%
\url{https://doi.org/10.14778/3654621.3654629}
\showDOI{\tempurl}


\bibitem[Schwab(2025)]%
        {laddering}
\bibfield{author}{\bibinfo{person}{Charles Schwab}.}
  \bibinfo{year}{2025}\natexlab{}.
\newblock \bibinfo{booktitle}{\emph{{Bond Ladders}}}.
\newblock
\urldef\tempurl%
\url{https://www.schwab.com/fixed-income/bond-ladders}
\showURL{%
Retrieved March 9, 2025 from \tempurl}


\bibitem[Snowflake(2024a)]%
        {snowadaptnet}
\bibfield{author}{\bibinfo{person}{Snowflake}.}
  \bibinfo{year}{2024}\natexlab{a}.
\newblock \bibinfo{booktitle}{\emph{{Adaptive Network Optimizations for Faster
  Query Performance}}}.
\newblock
\urldef\tempurl%
\url{https://www.snowflake.com/engineering-blog/adaptive-network-optimizations-faster-query-performance/}
\showURL{%
Retrieved October 17, 2024 from \tempurl}


\bibitem[Snowflake(2024b)]%
        {spi27blog}
\bibfield{author}{\bibinfo{person}{Snowflake}.}
  \bibinfo{year}{2024}\natexlab{b}.
\newblock \bibinfo{booktitle}{\emph{{Snowflake Improves Performance by 27\%,
  According to the Snowflake Performance Index}}}.
\newblock
\urldef\tempurl%
\url{https://www.snowflake.com/en/blog/performance-index-27-percent-improvement/}
\showURL{%
Retrieved October 17, 2024 from \tempurl}


\bibitem[Snowflake(2024c)]%
        {snowspi}
\bibfield{author}{\bibinfo{person}{Snowflake}.}
  \bibinfo{year}{2024}\natexlab{c}.
\newblock \bibinfo{booktitle}{\emph{{Snowflake Performance Index}}}.
\newblock
\urldef\tempurl%
\url{https://www.snowflake.com/en/data-cloud/pricing/performance-index/}
\showURL{%
Retrieved October 17, 2024 from \tempurl}


\bibitem[Sriraman et~al\mbox{.}(2019)]%
        {softsku}
\bibfield{author}{\bibinfo{person}{Akshitha Sriraman},
  \bibinfo{person}{Abhishek Dhanotia}, {and} \bibinfo{person}{Thomas~F.
  Wenisch}.} \bibinfo{year}{2019}\natexlab{}.
\newblock \showarticletitle{SoftSKU: Optimizing Server Architectures for
  Microservice Diversity @Scale}. In \bibinfo{booktitle}{\emph{2019 ACM/IEEE
  46th Annual International Symposium on Computer Architecture (ISCA)}}.
  \bibinfo{pages}{513--526}.
\newblock


\bibitem[Stokely et~al\mbox{.}(2025)]%
        {shavedicedata}
\bibfield{author}{\bibinfo{person}{Murray Stokely}, \bibinfo{person}{Jack
  Peele}, \bibinfo{person}{Neel Nadgir}, {and} \bibinfo{person}{Orestis
  Kostakis}.} \bibinfo{year}{2025}\natexlab{}.
\newblock \bibinfo{booktitle}{\emph{Shaved Ice Compute Resource Commitment
  Dataset}}.
\newblock
\urldef\tempurl%
\url{https://doi.org/10.5281/zenodo.15015993}
\showDOI{\tempurl}
\newblock
\shownote{GitHub: https://github.com/Snowflake-Labs/shavedice-dataset}.


\bibitem[Stokely et~al\mbox{.}(2009)]%
        {resourcemarkets}
\bibfield{author}{\bibinfo{person}{Murray Stokely}, \bibinfo{person}{Jim
  Winget}, \bibinfo{person}{Ed Keyes}, \bibinfo{person}{Carrie Grimes}, {and}
  \bibinfo{person}{Benjamin Yolken}.} \bibinfo{year}{2009}\natexlab{}.
\newblock \showarticletitle{Using a Market Economy to Provision Compute
  Resources Across Planet-wide Clusters}. In \bibinfo{booktitle}{\emph{2009
  IEEE International Symposium on Parallel \& Distributed Processing}}. IEEE,
  \bibinfo{pages}{1--8}.
\newblock


\bibitem[Straesser et~al\mbox{.}(2022)]%
        {autoscaling}
\bibfield{author}{\bibinfo{person}{Martin Straesser}, \bibinfo{person}{Johannes
  Grohmann}, \bibinfo{person}{J\'{o}akim von Kistowski}, \bibinfo{person}{Simon
  Eismann}, \bibinfo{person}{Andr\'{e} Bauer}, {and} \bibinfo{person}{Samuel
  Kounev}.} \bibinfo{year}{2022}\natexlab{}.
\newblock \showarticletitle{Why Is It Not Solved Yet? Challenges for
  Production-Ready Autoscaling}. In \bibinfo{booktitle}{\emph{Proceedings of
  the 2022 ACM/SPEC on International Conference on Performance Engineering}}
  (Beijing, China) \emph{(\bibinfo{series}{ICPE '22})}.
  \bibinfo{publisher}{Association for Computing Machinery},
  \bibinfo{address}{New York, NY, USA}, \bibinfo{pages}{105–115}.
\newblock
\showISBNx{9781450391436}
\urldef\tempurl%
\url{https://doi.org/10.1145/3489525.3511680}
\showDOI{\tempurl}


\bibitem[Sukprasert et~al\mbox{.}(2024)]%
        {temporalshifting}
\bibfield{author}{\bibinfo{person}{Thanathorn Sukprasert},
  \bibinfo{person}{Abel Souza}, \bibinfo{person}{Noman Bashir},
  \bibinfo{person}{David Irwin}, {and} \bibinfo{person}{Prashant Shenoy}.}
  \bibinfo{year}{2024}\natexlab{}.
\newblock \showarticletitle{On the Limitations of Carbon-Aware Temporal and
  Spatial Workload Shifting in the Cloud}. In
  \bibinfo{booktitle}{\emph{Proceedings of the Nineteenth European Conference
  on Computer Systems}} (Athens, Greece) \emph{(\bibinfo{series}{EuroSys
  '24})}. \bibinfo{publisher}{Association for Computing Machinery},
  \bibinfo{address}{New York, NY, USA}, \bibinfo{pages}{924–941}.
\newblock
\showISBNx{9798400704376}
\urldef\tempurl%
\url{https://doi.org/10.1145/3627703.3650079}
\showDOI{\tempurl}


\bibitem[Talluri et~al\mbox{.}(2019)]%
        {databricksworkload}
\bibfield{author}{\bibinfo{person}{Sacheendra Talluri}, \bibinfo{person}{Alicja
  \L{}uszczak}, \bibinfo{person}{Cristina~L. Abad}, {and}
  \bibinfo{person}{Alexandru Iosup}.} \bibinfo{year}{2019}\natexlab{}.
\newblock \showarticletitle{Characterization of a Big Data Storage Workload in
  the Cloud}. In \bibinfo{booktitle}{\emph{Proceedings of the 2019 ACM/SPEC
  International Conference on Performance Engineering}} (Mumbai, India)
  \emph{(\bibinfo{series}{ICPE '19})}. \bibinfo{publisher}{Association for
  Computing Machinery}, \bibinfo{address}{New York, NY, USA},
  \bibinfo{pages}{33–44}.
\newblock
\showISBNx{9781450362399}
\urldef\tempurl%
\url{https://doi.org/10.1145/3297663.3310302}
\showDOI{\tempurl}


\bibitem[Taylor and Letham(2018)]%
        {prophet}
\bibfield{author}{\bibinfo{person}{Sean~J Taylor} {and}
  \bibinfo{person}{Benjamin Letham}.} \bibinfo{year}{2018}\natexlab{}.
\newblock \showarticletitle{Forecasting at Scale}.
\newblock \bibinfo{journal}{\emph{The American Statistician}}
  \bibinfo{volume}{72}, \bibinfo{number}{1} (\bibinfo{year}{2018}),
  \bibinfo{pages}{37--45}.
\newblock


\bibitem[Thusoo et~al\mbox{.}(2010)]%
        {facebookdw}
\bibfield{author}{\bibinfo{person}{Ashish Thusoo}, \bibinfo{person}{Zheng
  Shao}, \bibinfo{person}{Suresh Anthony}, \bibinfo{person}{Dhruba Borthakur},
  \bibinfo{person}{Namit Jain}, \bibinfo{person}{Joydeep Sen~Sarma},
  \bibinfo{person}{Raghotham Murthy}, {and} \bibinfo{person}{Hao Liu}.}
  \bibinfo{year}{2010}\natexlab{}.
\newblock \showarticletitle{Data Warehousing and Analytics Infrastructure at
  Facebook}. In \bibinfo{booktitle}{\emph{Proceedings of the 2010 ACM SIGMOD
  International Conference on Management of data}}.
  \bibinfo{pages}{1013--1020}.
\newblock


\bibitem[Tseng et~al\mbox{.}(2017)]%
        {tseng2017dynamic}
\bibfield{author}{\bibinfo{person}{Fan-Hsun Tseng}, \bibinfo{person}{Xiaofei
  Wang}, \bibinfo{person}{Li-Der Chou}, \bibinfo{person}{Han-Chieh Chao}, {and}
  \bibinfo{person}{Victor~CM Leung}.} \bibinfo{year}{2017}\natexlab{}.
\newblock \showarticletitle{Dynamic Resource Prediction and Allocation for
  Cloud Data Center Using the Multiobjective Genetic Algorithm}.
\newblock \bibinfo{journal}{\emph{IEEE Systems Journal}} \bibinfo{volume}{12},
  \bibinfo{number}{2} (\bibinfo{year}{2017}), \bibinfo{pages}{1688--1699}.
\newblock


\bibitem[Vieira et~al\mbox{.}(2014)]%
        {reducecostsbrazil}
\bibfield{author}{\bibinfo{person}{Cristiano~C.A. Vieira},
  \bibinfo{person}{Luiz~F. Bittencourt}, {and} \bibinfo{person}{Edmundo~R.M.
  Madeira}.} \bibinfo{year}{2014}\natexlab{}.
\newblock \showarticletitle{Reducing Costs in Cloud Application Execution Using
  Redundancy-Based Scheduling}. In \bibinfo{booktitle}{\emph{2014 IEEE/ACM 7th
  International Conference on Utility and Cloud Computing}}.
  \bibinfo{pages}{117--126}.
\newblock
\urldef\tempurl%
\url{https://doi.org/10.1109/UCC.2014.20}
\showDOI{\tempurl}


\bibitem[Vuppalapati et~al\mbox{.}(2020)]%
        {snowflakeelasticquery}
\bibfield{author}{\bibinfo{person}{Midhul Vuppalapati}, \bibinfo{person}{Justin
  Miron}, \bibinfo{person}{Rachit Agarwal}, \bibinfo{person}{Dan Truong},
  \bibinfo{person}{Ashish Motivala}, {and} \bibinfo{person}{Thierry Cruanes}.}
  \bibinfo{year}{2020}\natexlab{}.
\newblock \showarticletitle{Building an Elastic Query Engine on Disaggregated
  Storage}. In \bibinfo{booktitle}{\emph{Proceedings of the 17th Usenix
  Conference on Networked Systems Design and Implementation}} (Santa Clara, CA,
  USA) \emph{(\bibinfo{series}{NSDI'20})}. \bibinfo{publisher}{USENIX
  Association}, \bibinfo{address}{USA}, \bibinfo{pages}{449–462}.
\newblock
\showISBNx{9781939133137}


\bibitem[Wang et~al\mbox{.}(2013)]%
        {toreserve}
\bibfield{author}{\bibinfo{person}{Wei Wang}, \bibinfo{person}{Baochun Li},
  {and} \bibinfo{person}{Ben Liang}.} \bibinfo{year}{2013}\natexlab{}.
\newblock \showarticletitle{To Reserve or Not to Reserve: Optimal Online
  {Multi-Instance} Acquisition in {IaaS} Clouds}. In
  \bibinfo{booktitle}{\emph{10th International Conference on Autonomic
  Computing (ICAC 13)}}. \bibinfo{publisher}{USENIX Association},
  \bibinfo{address}{San Jose, CA}, \bibinfo{pages}{13--22}.
\newblock
\showISBNx{978-1-931971-02-7}
\urldef\tempurl%
\url{https://www.usenix.org/conference/icac13/technical-sessions/presentation/wang_wei}
\showURL{%
\tempurl}


\bibitem[Wiesner et~al\mbox{.}(2021)]%
        {letswait}
\bibfield{author}{\bibinfo{person}{Philipp Wiesner}, \bibinfo{person}{Ilja
  Behnke}, \bibinfo{person}{Dominik Scheinert}, \bibinfo{person}{Kordian
  Gontarska}, {and} \bibinfo{person}{Lauritz Thamsen}.}
  \bibinfo{year}{2021}\natexlab{}.
\newblock \showarticletitle{Let's Wait Awhile: How Temporal Workload Shifting
  Can Reduce Carbon Emissions in the Cloud}. In
  \bibinfo{booktitle}{\emph{Proceedings of the 22nd International Middleware
  Conference}} (Qu\'{e}bec city, Canada) \emph{(\bibinfo{series}{Middleware
  '21})}. \bibinfo{publisher}{Association for Computing Machinery},
  \bibinfo{address}{New York, NY, USA}, \bibinfo{pages}{260–272}.
\newblock
\showISBNx{9781450385343}
\urldef\tempurl%
\url{https://doi.org/10.1145/3464298.3493399}
\showDOI{\tempurl}


\bibitem[Yan et~al\mbox{.}(2018)]%
        {snowtrail}
\bibfield{author}{\bibinfo{person}{Jiaqi Yan}, \bibinfo{person}{Qiuye Jin},
  \bibinfo{person}{Shrainik Jain}, \bibinfo{person}{Stratis~D. Viglas}, {and}
  \bibinfo{person}{Allison Lee}.} \bibinfo{year}{2018}\natexlab{}.
\newblock \showarticletitle{Snowtrail: Testing with Production Queries on a
  Cloud Database}. In \bibinfo{booktitle}{\emph{Proceedings of the Workshop on
  Testing Database Systems}} (Houston, TX, USA) \emph{(\bibinfo{series}{DBTest
  '18})}. \bibinfo{publisher}{Association for Computing Machinery},
  \bibinfo{address}{New York, NY, USA}, Article \bibinfo{articleno}{4},
  \bibinfo{numpages}{6}~pages.
\newblock
\showISBNx{9781450358262}
\urldef\tempurl%
\url{https://doi.org/10.1145/3209950.3209958}
\showDOI{\tempurl}


\bibitem[Yao et~al\mbox{.}(2021)]%
        {mlvmprovisioning}
\bibfield{author}{\bibinfo{person}{R. Yao}, \bibinfo{person}{C. Luo},
  \bibinfo{person}{B. Qiao}, \bibinfo{person}{Q. Lin}, \bibinfo{person}{T.
  Tran}, \bibinfo{person}{G. Shafriri}, \bibinfo{person}{Y. Dang},
  \bibinfo{person}{R. Ghelman}, \bibinfo{person}{P. Goyal}, \bibinfo{person}{E.
  Cortez}, \bibinfo{person}{D. Howlader}, \bibinfo{person}{S. Rewaskar},
  \bibinfo{person}{M. Chintalapati}, {and} \bibinfo{person}{D. Zhang}.}
  \bibinfo{year}{2021}\natexlab{}.
\newblock \showarticletitle{Infusing ML into VM Provisioning in Cloud}. In
  \bibinfo{booktitle}{\emph{2021 IEEE/ACM International Workshop on Cloud
  Intelligence (CloudIntelligence)}}. \bibinfo{publisher}{IEEE Computer
  Society}, \bibinfo{address}{Los Alamitos, CA, USA}, \bibinfo{pages}{44--45}.
\newblock
\urldef\tempurl%
\url{https://doi.org/10.1109/CloudIntelligence52565.2021.00018}
\showDOI{\tempurl}


\end{thebibliography}



\end{document}